\documentclass[aps,reprint]{revtex4-2}
\usepackage[english]{babel}
\usepackage[utf8]{inputenc}
\usepackage{xcolor}
\usepackage{physics}
\usepackage[colorinlistoftodos, color=green!40, prependcaption]{todonotes}
\usepackage{amsthm,amsmath,amssymb,bm}
\usepackage{array}
\usepackage{mathtools}
\usepackage{physics}
\usepackage{graphicx}
\usepackage[left=15mm,right=15mm,top=20mm,columnsep=15pt]{geometry} 
\usepackage{adjustbox}
\usepackage{placeins}
\usepackage[T1]{fontenc}
\usepackage{lipsum}
\usepackage{csquotes}
\usepackage{float}
\usepackage[pdftex, pdftitle={Article}, pdfauthor={Author}]{hyperref} 

\def\bs#1{\boldsymbol{#1}}
\def\rmd{{\rm d}}
\def\rmi{\mathrm{i}}

\def\diff#1#2{{{\rmd #1} \over {\rmd#2}}}

\usepackage[inline]{trackchanges}
\addeditor{AO}
\addeditor{LK}

\begin{document}

\title{Phase-field modeling of colloid-polymer mixtures in microgravity}

 \author{Lauren Barnes$^{1}$, Boris Khusid$^{2}$, Lou Kondic$^{1}$, William V. Meyer$^{3}$, Anand U. Oza$^{1}$}

 \email[Correspondence email addresses: ]{oza@njit.edu}

 \affiliation{$^{1}$Department of Mathematical Sciences \& Center for Applied Mathematics and Statistics, New Jersey Institute of Technology, Newark, New Jersey 07102, USA}
 
 \affiliation{$^{2}$Otto H. York Department of Chemical and Materials Engineering, New Jersey Institute of Technology, Newark, New Jersey 07102, USA}
  
 \affiliation{$^{3}$USRA at NASA Glenn Research Center, Cleveland, Ohio 44135, USA}  

\date{\today} 

\begin{abstract}

Colloid-polymer mixtures are an archetype for modeling phase transition processes, as they a exhibit low-density gas phase, high-density crystalline phase and an intervening liquid phase. While their equilibrium behavior has been studied extensively, the role of hydrodynamics in driving their phase separation is not yet understood. We present a theoretical model that describes hydrodynamic interactions in colloid-polymer mixtures in a microgravity environment. Our phase-field model consists of the Cahn-Hilliard equation, which describes phase separation processes in multicomponent mixtures, coupled with the Stokes equation for viscous fluid flow. We account for the dependence of the suspension viscosity on the colloid concentration, and the so-called Korteweg stresses that arise at the interfaces of colloidal phases. We process video microscopy images from NASA's Binary Colloid Alloy Test (BCAT) experiments, which were performed on the International Space Station. While terrestrial experiments would be dominated by gravitational forces and buoyancy-driven flows, the microgravity environment of the BCAT experiments allows for the visualization of phase separation by low interfacial tension, and thus enables a quantitative comparison between experiment and our model predictions. 
\end{abstract}

\maketitle

\section{Introduction}

The phase transition behavior of colloidal suspensions exhibits striking similarities to that of atomic-scale systems, providing insight into systems that are otherwise difficult to observe~\cite{Poon2002,Anderson2002}. Because the colloid particles are macroscopic, their behavior is readily observable via optical video microscopy and small-angle light scattering~\cite{Pusey1994}. In addition to scientific curiosity, industrial applications of colloidal systems include materials engineering, chemical processing and the manufacturing of pharmaceutical products. 

A deeper understanding of the structure and dynamics of colloidal suspensions and gels could also enable self-assembly of new materials~\cite{Manoharan2015}, which are potentially useful in space~\cite{Decadal2023}. Moreover, their structure and dynamics in a microgravity environment could cause them to acquire new microstructures and physical properties, and thus lead to the manufacturing of new soft materials in low-Earth-orbit. On a scientific level, experiments in space allow for the effects of gravitational forces and buoyancy-driven flows to be suppressed, allowing for other relatively weak driving forces to be probed. The knowledge gained from such experiments could have Earth-bound applications, such as the stabilizing colloidal solutions, gels and foams used by consumers~\cite{Decadal2023}.

While the extensive literature on crystallization in colloidal suspensions has been reviewed elsewhere~\cite{Palberg1999,Palberg2014}, we here give a brief historical introduction to the subject for the sake of completeness. Seminal computer simulations~\cite{Alder1957}  demonstrated that a system of hard spheres could exhibit a fluid-to-crystal phase transition. This prediction was confirmed in experiments~\cite{Pusey1986}, where a progression from colloidal fluid, to fluid and crystal phases in coexistence, to full crystallization was observed as the colloid concentration was increased progressively. The phase transitions are now understood to be driven by entropy alone. 
Those experiments also suggested that, as the colloid concentration was increased progressively, the crystalline phase transitioned to a glass-like state, wherein the colloidal particles were jammed in a disordered fashion. This tantalizing discovery motivated microgravity experiments conducted on the Space Shuttle (STS-73)~\cite{Zhu1997}: while a phase transition from liquid to crystal was observed, the glassy state was not, indicating that it may be an artifact of gravitational effects. The crystalline state observed in those experiments has been shown to have the face-centered cubic (fcc) geometry~\cite{Cheng2001,Lei2024}. 


While colloidal suspensions are interesting in their own right, the addition of non-adsorbing polymer to a colloidal suspension makes the phase behavior even richer~\cite{Gast1998}. This is due to the short-range attractive depletion forces that arise when two solid bodies (e.g. colloids) are immersed in a polymer solution. The substantial literature on colloid-polymer suspensions has been reviewed elsewhere~\cite{Poon2002,Tuinier2003}, some of which we highlight in what follows. Gast {\it et al.}~\cite{Gast1983} and Lekkerkerker {\it et al.}~\cite{Lekkerkerker1992} built on theories of the depletion interaction~\cite{Asakura1954,Vrij1977} in order to investigate the phase behavior of colloid-polymer suspensions. They found that, for polymers small relative to the colloid, a colloid-polymer mixture admits two equilibrium phases, a fluid and a crystal, analogous to the colloid-only system described in the preceding paragraph. Above a critical polymer size, the system was found to admit three phases: a gas, a crystal, and an intermediate liquid phase. These predictions were confirmed in terrestrial experiments~\cite{Ilett1995} on suspensions of polymethylmethacrylate (PMMA) spheres and the polymer polystyrene in the solvent cis-decalin. The phase diagram delineating the dependence of the equilibrium phase on colloid and polymer concentrations has been determined experimentally~\cite{Calderon1993,Poon1993,Pusey1993,deHoog1999}. Subsequent experiments investigated the structure of the suspensions near the triple point, a regime in which the gas, liquid and crystalline phase coexist~\cite{Moussaid1999}. The kinetics of phase separation was also studied experimentally~\cite{Poon1999,Renth2001}, where it was found that the three phases can emerge via distinct pathways depending on the colloid volume fraction and polymer concentration. Non-equilibrium aggregation of colloids and subsequent gelation can occur for sufficiently strong depletion forces, as generated by increasing the polymer concentration~\cite{Poon1997,Zaccarelli2007}.

Despite the insights provided by the aforementioned experiments, it has been recognized that gravitational forces can qualitatively affect phase separation~\cite{Verhaegh1996} and gel formation~\cite{Segre2001} in colloid-polymer mixtures, as colloids often sediment on a shorter timescale than that required to observe complete phase separation. Indeed, terrestrial experiments are typically dominated by gravitational forces and buoyancy-induced flows; these effects obscure the details of dynamics driven by low interfacial tension, making microgravity experiments valuable platforms for investigating colloidal phase separation and testing theoretical models~\cite{Decadal2023}. Experiments were thus conducted in microgravity on the International Space Station~\cite{Bailey2007}, where it was observed that spinodal decomposition at early stages of the experiment eventually gave way to fluid-driven coarsening at long timescales. Other microgravity experiments~\cite{Sabin2012} revealed that, in certain parameter regimes, gas-liquid phase separation could be arrested due to the formation of a crystal gel consisting of crystalline strands that run through the entire sample. 

A number of prior works have studied colloid-polymer mixtures using simulations that neglect hydrodynamic effects, for example, using Brownian dynamics~\cite{Soga1998,dArjuzon2003,Cerda2004}, molecular dynamics~\cite{Costa2002}, Monte Carlo simulations~\cite{Meijer1991,Meijer1994,Dijkstra1999} and dynamic density functional theory~\cite{Schmidt2000,Schmidt2002}, to name a few. Phenomenological Cahn-Hilliard theories, also without hydrodynamics, have been used to rationalize the kinetic pathways that drive phase separation in colloid-polymer mixtures~\cite{Evans2001}, and to understand the role of an intermediate liquid phase in unbinding a gas-solid interface~\cite{Evans1997_1,Evans1997_2,Evans1997_3}. However, a discrepancy between theory and experiment in a recent study of a two-dimensional colloid-polymer mixture was attributed to hydrodynamic screening, the effect of which was neglected in the theory~\cite{Griffiths2021}. Moreover, lattice-Boltzmann simulations of a colloid-polymer mixture showed that, while the final gel structure is not affected by hydrodynamics, the speed of gelation is~\cite{Graaf2019}. Simulations of colloidal suspensions (without polymer) using the ``fluid particle dynamics" method~\cite{Tanaka2000b,Furukawa2018}, in which colloids are modeled as high-viscosity fluids, also revealed the crucial role of hydrodynamics in mediating phase separation~\cite{Tateno2019}. 

The goal of this paper is to construct and simulate a phase-field model~\cite{Anderson1998,Singer-Loginova2008} that describes phase separation via spinodal decomposition in colloid-polymer mixtures. Our model incorporates hydrodynamic effects relevant to such mixtures, specifically, by accounting for the increase in suspension viscosity with colloid concentration, and incorporating the so-called Korteweg stresses that arise due to gradients in the colloid concentration. While Korteweg stresses have recently been incorporated in volume-of-fluid simulations of colloidal dispersions~\cite{Gowda2019,Gowda2021}, to our knowledge they have not been applied to the study of colloid-polymer mixtures. We compare our theoretical predictions with NASA's Binary Colloid Alloy Test experiments, which were conducted in microgravity on the International Space Station. Previous analysis of spinodal decomposition in these experiments was based primarily on data collected via small-angle light scattering~\cite{Bailey2007}. This analysis provides a good description of relatively small-scale cluster growth, 
but is generally less useful for investigating the large-scale structuring that occurs. We thus processed the video microscopy images from the experiments, which are available on NASA's Physical Sciences Informatics (PSI) database, and used the images to quantitatively characterize the coarsening process. 

The paper is organized as follows. In \S\ref{Sec:Image} we describe the algorithm we used to process the video microscopy images from the BCAT experiments, and to obtain a quantitative description of the coarsening rate. We construct the phase-field model in \S\ref{Sec:Model}, and describe the algorithm we use to solve it in \S\ref{Sec:Numerics}. Qualitative features observed in simulations of the phase-field model are described in \S\ref{Sec:results}. In \S\ref{Sec:Theoryvsexp} we compare the coarsening rates predicted by our model against those obtained in the BCAT experiments, with and without the consideration of hydrodynamics. Conclusions and future directions are presented in \S\ref{Sec:Discussion}.

\section{Processing Images from BCAT Experiments}\label{Sec:Image}

\begin{table*}[t]
\centering
{\setlength{\extrarowheight}{6pt}
\setlength{\tabcolsep}{10pt}
\begin{tabular}{|c||c|c|c|c|c|c|c|c|}
\hline   Sample  & Year & $\phi_0$ & $a$ (nm) & $\rho$ (mg/mL) & $\xi = \delta/a$ & $(4/3)\pi\delta^3 n_{\text{R}}$ \\ \hline \hline
BCAT-3 Sample 1 & 2006 & 0.2273 & 97 & 0.814 &  1.24 & 24.58 \\ \hline
BCAT-3 Sample 2 & 2006 & 0.22 & 97 & 0.781 &  1.24 & 18.57 \\ \hline
BCAT-3 Sample 4 & 2008 & 0.21 & 97 & 0.737 &  1.24 & 12.83 \\ \hline
BCAT-3 Sample 6 & 2006 & 0.2112 & 97 & 0.742  & 1.24 & 13.40 \\ \hline
BCAT-4 Sample 1  & 2008 & 0.2237 & 97 & 0.797  & 1.24 & 21.37 \\ \hline
BCAT-4 Sample 2  & 2013 & 0.2173 & 97 & 0.770  & 1.24 & 16.80 \\ \hline
BCAT-4 Sample 3 & 2013 & 0.2151 & 97 & 0.760  & 1.24 & 15.48 \\ \hline
BCAT-5 Sample 4  & 2011 & 0.223 & 216 & 0.797  & 0.56 & 0.92 \\ \hline
BCAT-5 Sample 5 & 2011 & 0.2173 & 216 & 0.770  & 0.56 & 0.85 \\ \hline
BCAT-5 Sample 6  & 2009 & 0.29 & 229 & 0.72 & 0.52 & 1.23 \\ \hline
BCAT-5 Sample 7  & 2010 & 0.24 & 229 & 0.88 & 0.52 & 0.99 \\ \hline
BCAT-5 Sample 8 & 2010 & 0.35 & 229 & 0.55 & 0.52 & 1.81 \\ \hline
\end{tabular}
}
\caption{Table of the colloid volume fraction $\phi_0$, colloid radius $a$ and polymer mass concentration $\rho$, and other derived quantities corresponding to the BCAT experiments. The solvent composition is 47:53 decalin/tetralin for the BCAT-3 and BCAT-4 experiments, and 45:55 for the BCAT-5 experiments. The temperature is $T = 295$ K and the polymer radius of gyration is $\delta = 120$ nm. The second column of the table shows the year the images were taken, as reported on the NASA PSI database. The PIs for the BCAT-3 experiments were D. Weitz, P. N. Pusey, A. G. Yodh, P. M. Chaikin and W. B. Russel. The PI for the BCAT-4 Samples 1-3 and BCAT-5 Samples 4-5 was D. Weitz, and the co-I was P. Lu. The PI for BCAT-5 Samples 6-8 was B. Frisken, and the co-I was A. Bailey.}
\label{tab:param}
\end{table*}

NASA's PSI database contains plentiful information about the BCAT experiments~\cite{NASAPSI_Images}. The four BCAT experiments conducted, BCAT-3, 4, 5 and 6, each consist of ten samples at room temperature, $T= 295$ K~\cite{NASAPSI_SciDocs}. In all of these samples, the colloids are polymethyl-methacrylate (PMMA) spheres, the solvent is a mixture of decalin and tetralin, and the polymer is polystyrene~\cite{NASAPSI_ExpData,NASAPSI_SciDocs} with a radius of gyration $\delta = 120$ nm. Photographs of the colloid-polymer mixtures were taken just after mixing and then approximately every thirty minutes after that. Each sample has different values for the initial (homogenized) colloid volume fraction $\phi_0$, polymer mass concentration  $\rho$ and colloid radius $a$. Several of the samples were aimed toward studying other phenomena, such as seeded crystal growth, and contained colloids of varying sizes; others had only colloids and no polymer. Those samples are not considered herein, as we are interested in the phase behavior of colloid-polymer suspensions. As stated in the BCAT-6 Final Summary Report, available on the PSI database, no phase separation was observed in the BCAT-6 samples, so we focused on BCAT-3, BCAT-4 and BCAT-5. There were also a few samples that we excluded because the images were of low quality and not amenable to analysis. The relevant samples for the purposes of this paper are Samples 1, 2, 4 and 6 of BCAT-3, Samples 1-3 of BCAT-4, and Samples 4-8 of BCAT-5.  The dimensions of the visible volume of the sample cells are $4\times10\times20$ mm~\cite{Sabin2012}. The relevant parameter values for each of the samples are given in Table~\ref{tab:param}. 
The parameter $n_{\mathrm{R}}$ in the last column of the table is related to the polymer concentration and will be defined in \S\ref{Sec:Model}.  In this paper, we focus on BCAT-5, since for this experiment the ratio of the polymer radius of gyration $\delta$ to the colloid radius $a$ is less than unity, $\xi = \delta/a<1$, which is the regime in which the theory we develop in \S\ref{Sec:Model} is expected to be valid~\cite{Lekkerkerker1992}.

\begin{figure*}[hbtp]
    \centering
    \includegraphics[width=1\textwidth]{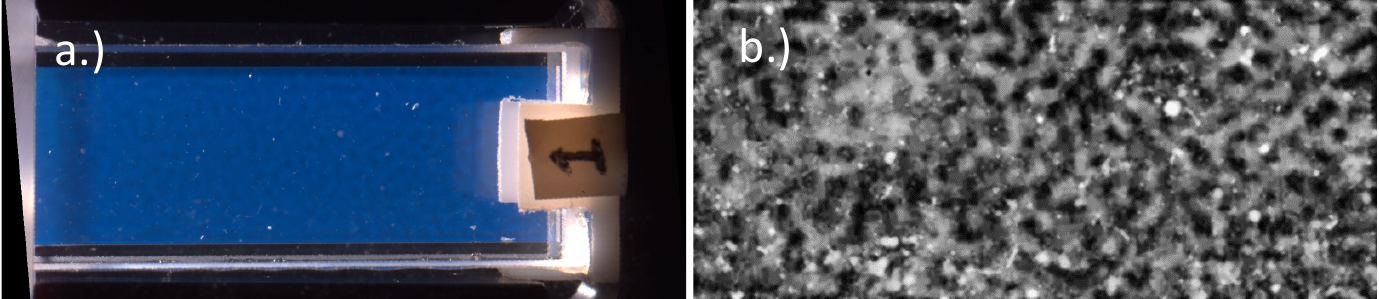}
    \caption{Image processing algorithm applied to an image of BCAT-3 Sample 1, as described in \S\ref{Sec:Image}. The raw and enhanced images are shown in panels (a) and (b), respectively. The raw image in panel (a) was obtained from the NASA PSI database. In panel (b), light (dark) areas indicate colloid-rich (poor) regions. The height of the sample domain is approximately 10 mm.}
    \label{fig:ImageEnhance}
\end{figure*}

\begin{figure*}
    \centering
    \includegraphics[width=1\textwidth]{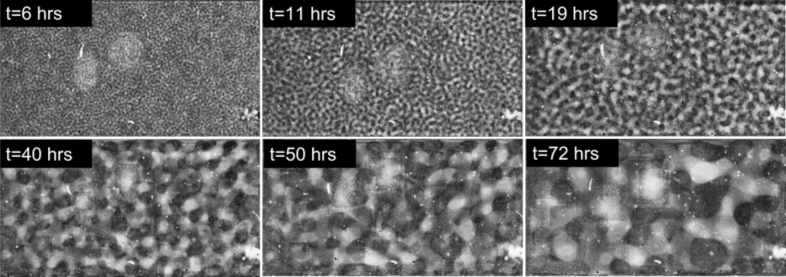}
    \caption{Time evolution of BCAT-5 Sample 4. The parameters are listed in Table~\ref{tab:param}. The time after mixing (in hours) is labeled at the top of each plot. The height of the sample domains pictured in these images is approximately 10 mm.}
    \label{fig:BCAT5-4}
\end{figure*}

\begin{figure*}
    \centering
    \includegraphics[width=1\textwidth]{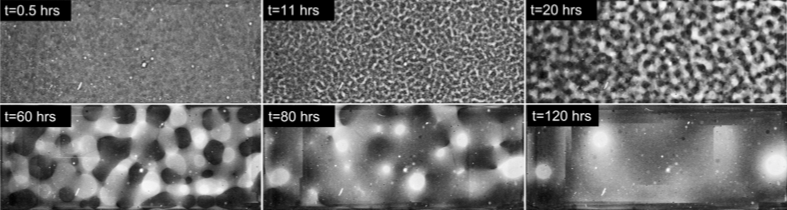}
    \caption{Time evolution of BCAT-5 Sample 5. The parameters are listed in Table~\ref{tab:param}.}
    \label{fig:BCAT5-5}
\end{figure*}

\begin{figure*}
    \centering
    \includegraphics[width=1\textwidth]{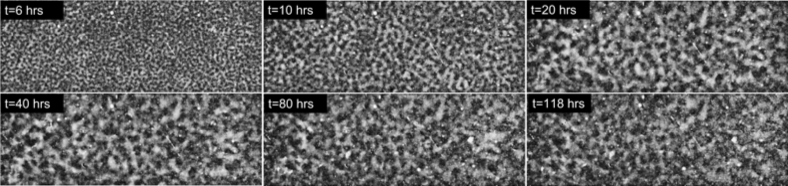}
    \caption{Time evolution of BCAT-5 Sample 6. The parameters are listed in Table~\ref{tab:param}.}
    \label{fig:BCAT5-6}
\end{figure*}

\begin{figure*}
    \centering
    \includegraphics[width=1\textwidth]{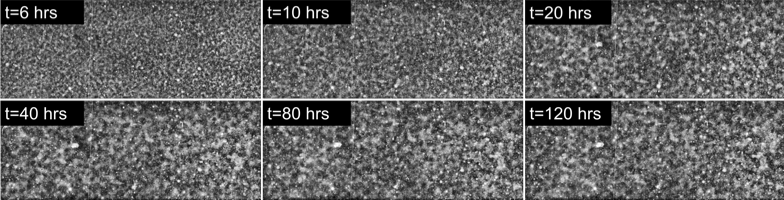}
    \caption{Time evolution of BCAT-5 Sample 7. The parameters are listed in Table~\ref{tab:param}.}
    \label{fig:BCAT5-7}
\end{figure*}

\begin{figure*}
    \centering
    \includegraphics[width=1\textwidth]{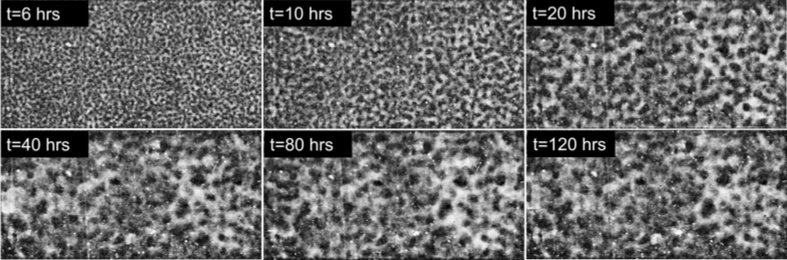}
    \caption{Time evolution of BCAT-5 Sample 8. The parameters are listed in Table~\ref{tab:param}.}
    \label{fig:BCAT5-8}
\end{figure*}

Our first step was to process the raw images from NASA's PSI database and make them conducive to quantitative characterization. Figure~\ref{fig:ImageEnhance} shows a BCAT image before and after processing. We developed an algorithm in MATLAB to first straighten each image and then crop it to eliminate the dark background, the frame of the cell containing the sample, and the sample number label, thus leaving only the sample domain. We then used the software ImageJ to adjust the image's brightness and contrast, subtract the background, and then enhance the contrast again. Finally, the image was smoothed and despeckled in ImageJ, producing the finished product. Additional implementation details are provided in Ref.~\cite[Chapter 4.2]{Barnes_Thesis}. Figures~\ref{fig:BCAT5-4}-\ref{fig:BCAT5-8} show time series of the enhanced images corresponding to BCAT-5 Samples 4 through 8. The corresponding images for the BCAT-3 and BCAT-4 experiments are shown in Figs.~\ref{fig:BCAT3-1}-\ref{fig:BCAT4-3} of Appendix~\ref{Sec:Appendix}.

As is evident from Figures~\ref{fig:BCAT5-4}-\ref{fig:BCAT5-8} and Appendix~\ref{Sec:Appendix} Figures~\ref{fig:BCAT3-1}-\ref{fig:BCAT4-3}, the samples exhibit similar qualitative behavior: at early times, the initially-homogenized mixture begins to separate into small domains of colloid-poor and colloid-rich phases. Then, these domains grow in size over a timescale of hours. To quantify the growth of the phase domains in time, we computed the characteristic domain length $\lambda_a$ for every image. Specifically, we computed the spatial autocorrelation function $C(\bs{x})$, then averaged $C$ azimuthally 
using the trapezoidal rule, which yields the 1D autocorrelation function $\tilde{C}(r)$. The characteristic length scale $\lambda_a$ for a given image is found by selecting the first maximum in the curve $\tilde{C}(r)$; this value is a measure of the typical distance between colloid-rich (or colloid-poor) domains. The resulting plots of the time evolution of $\lambda_a(t)$ are shown for BCAT-5 in Fig.~\ref{fig:BCAT5Auto}, and for BCAT-3 and BCAT-4 in Appendix~\ref{Sec:Appendix} Fig.~\ref{fig:BCAT34Auto}. 

\begin{figure}[hbtp]
    \centering
    \includegraphics[width=1\columnwidth]{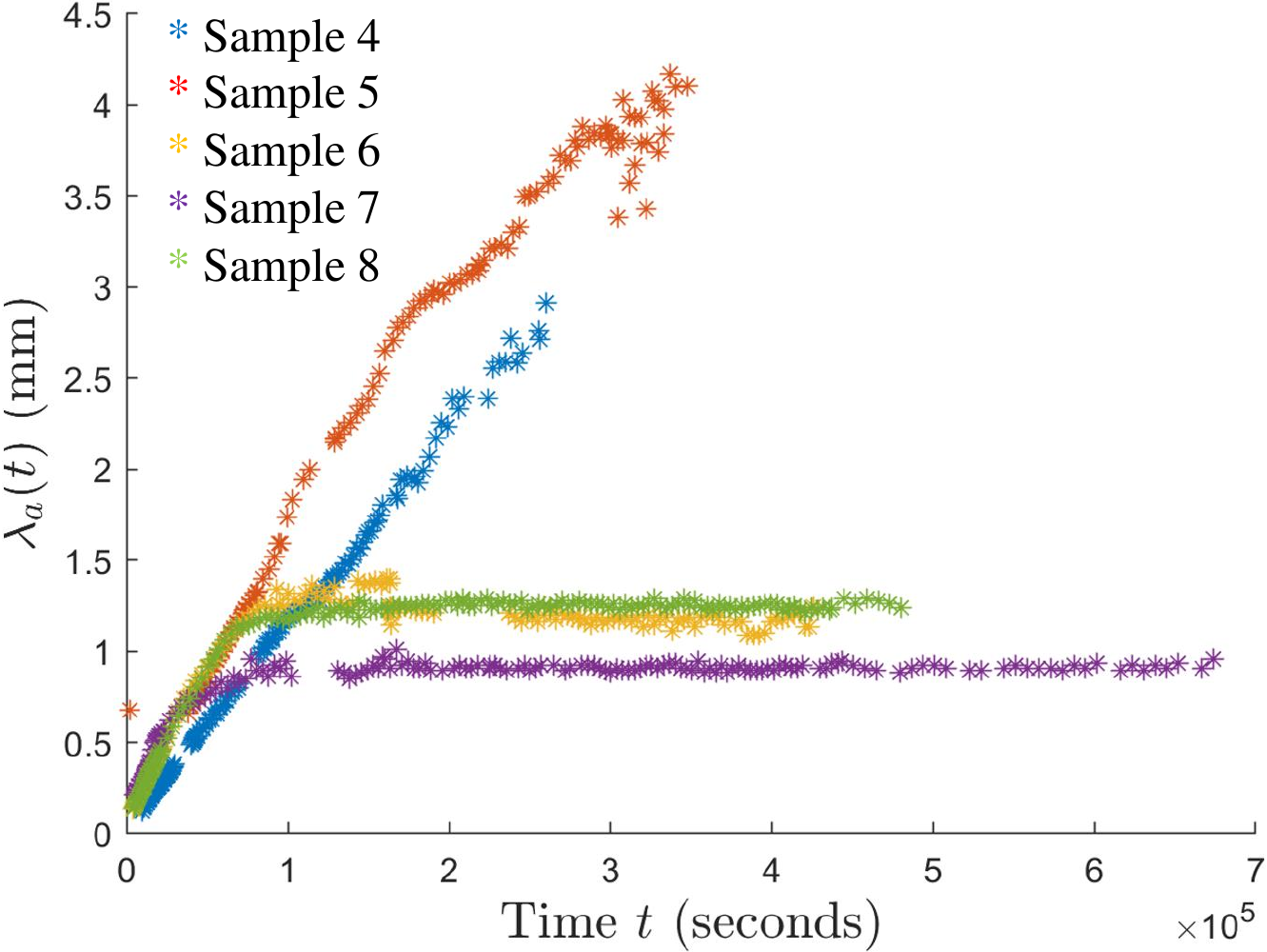}
    \caption{Time evolution of the characteristic length scale $\lambda_a(t)$ for the samples in the BCAT-5 experiments.}
    \label{fig:BCAT5Auto}
\end{figure}

We observe that $\lambda_a$ increases more rapidly for BCAT-5 Samples 4 and 5 (blue and red points in Fig.~\ref{fig:BCAT5Auto}) than for any of the BCAT-3 or BCAT-4 samples (Appendix~\ref{Sec:Appendix} Fig.~\ref{fig:BCAT34Auto}), despite the fact that the samples have similar colloid volume fractions $\phi_0$ and polymer concentrations $\rho$ (Table~\ref{tab:param}). 
Domain growth proceeds roughly linearly in time in BCAT-5 Samples 4 and 5, similarly to the samples in BCAT-3 and BCAT-4. 
However, the curves for BCAT-5 Samples 6, 7, and 8 (yellow, purple and green points in Fig.~\ref{fig:BCAT5Auto}) exhibit qualitatively different behavior from any of the others: the curves increase roughly linearly at first but then abruptly flatten out and remain roughly constant, indicating that the phase domain coarsening has stopped. The Sample 7 curve (purple) is the first to begin flattening out after approximately $13$ hours, followed by Samples 6 (yellow) and 8 (green), which cease to exhibit linear growth at around the same time as each other ($t\approx20$ hours.) These results are consistent with those of Sabin {\it et al.}~\cite{Sabin2012}, who reported the formation of crystallites having an arresting effect on the phase separation in the same experiments. We note that the values $\lambda_a(t)$ for BCAT-5 Samples 6, 7 and 8 agree well with those reported in Ref.~\cite{Sabin2012}, which serves as a validation of our image processing algorithm.

\section{Model development}\label{Sec:Model}

We here develop a model for describing phase separation in colloid-polymer mixtures. Our model consists of the Cahn-Hilliard equation, coupled with the incompressible Stokes equation for low-Reynolds number flow~\cite{Jacqmin1999}:
\begin{subequations}\label{CHStokes}
\begin{align}
&\left(\partial_t+\bs{u}\cdot\bs{\nabla}\right)\phi=\bs{\nabla}\cdot\left[\Gamma(\phi)\bs{\nabla}\left(\mu(\phi)-k_{\text{B}}T\ell^2\Delta\phi\right)\right],\label{CH} \\
&\bs{\nabla}\cdot\Big[\eta(\phi)\Big(\nabla\bs{u} + (\nabla\bs{u})^{\text{T}}\Big)\Big]=\bs{\nabla}p+\kappa\frac{k_{\text{B}}T\ell^2}{V_{\text{c}}}\bs{\nabla}\cdot\left(\bs{\nabla}\phi\bs{\nabla}\phi\right),\label{Stokes} \\
& \bs{\nabla}\cdot\bs{u}=0.\label{Incomp}
\end{align}
\end{subequations}
The Cahn-Hilliard equation~\eqref{CH} describes the dynamics of the colloid volume fraction $\phi(\bs{x},t)$. Here, $\Gamma(\phi)$ is the mobility, $\mu(\phi)$ the colloid chemical potential, $k_{\text{B}}$ Boltzmann's constant, $T$ the temperature and $\ell$ a free parameter characterizing the width of the interface between phases. We proceed by describing in turn the forms of $\Gamma(\phi)$ and $\mu(\phi)$.

The mobility $\Gamma(\phi)$ describes the extent to which the colloid particles are free to move about. 
We use the expression given by van Megen {\it et al.}~\cite{Megen1998}, which was also used in a prior numerical study on crystallite growth in a colloidal suspension~\cite{Lam2020}: 
\begin{align}
\Gamma_+(\phi)=\Gamma_0\left(1-\frac{\phi}{0.57}\right)^{2.6}.\label{GammaFunc}
\end{align} 
The factor $\Gamma_0$ is obtained from the Stokes-Einstein relation and has the value $\Gamma_0=1/(6\pi\eta_0 a)$, where $a$ is the colloid radius and $\eta_0$ the viscosity of the solvent in the absence of colloids. The critical value $\phi=0.57$ is the value at which the glass transition occurs, and the colloids are effectively no longer able to move freely~\cite{Megen1998}. Although negative values of $\phi$ are unphysical, Eq.~\eqref{CHStokes} is a phase-field model and so must be well-defined for {\it all} values of $\phi$. Thus, for the purposes of our simulations, we define $\Gamma$ as
\begin{align}
\Gamma(\phi) = \begin{cases} \Gamma_0 &\text{if }\phi\leq 0, \\ p_{\Gamma}(\phi) &\text{if }0 < \phi\leq 0.01, \\\Gamma_+(\phi) &\text{if }0.01 < \phi < 0.57, \\ 0&\text{if }\phi\geq 0.57,\end{cases}\label{GammaDef}
\end{align}
where $p_{\Gamma}(\phi)$ is a seventh-order polynomial spline interpolant for  $0 < \phi\leq 0.01$, defined in such a way that $\Gamma$ has three continuous derivatives. The cutoff value $\phi=0.01$ was chosen so that the polynomial interpolant would have coefficients of moderate size and also have minimal influence on the simulations. 

We define the colloid chemical potential $\mu(\phi)$ to be a piecewise-defined function:
\begin{align}
\mu(\phi) = \begin{cases} p_{\mu,1}(\phi) &\text{if }\phi\leq 0.01, \\ \mu_{\text{F}}(\phi) &\text{if }0.01 < \phi\leq 0.495, \\ p_{\mu,2}(\phi) &\text{if }0.495 < \phi < 0.54, \\ \mu_{\text{S}}(\phi) &\text{if }\phi\geq 0.54.\end{cases}
\end{align}
Here, we follow Lekkerkerker {\it et al.}~\cite{Lekkerkerker1992} and use the chemical potentials $\mu_{\text{F}}(\phi)$ of the colloid-fluid phase for $\phi\leq 0.495$ and $\mu_{\text{S}}(\phi)$ of the colloid-crystal phase for $\phi\geq 0.54$. The values $\phi_1 = 0.494$ and $\phi_2 = 0.545$ are the coexistence compositions in the absence of polymer; for values $\phi_1<\phi<\phi_2$, the system is in the fluid-solid coexistence regime~\cite{Hoover1968}, in which we define $\mu(\phi)$ by a seventh-order polynomial function $p_{\mu,2}(\phi)$ that connects the fluid and crystal chemical potentials so that $\mu(\phi)$ is three-times differentiable. We also use a fourth-order polynomial $p_{\mu,1}(\phi)$ to extend the definition of the chemical potential to $\phi < 0$, while maintaining the requirement that $\mu(\phi)$ be three-times differentiable.  The chemical potentials $\mu_{\text{F}}(\phi)$ and $\mu_{\text{S}}(\phi)$ are defined by the expression
\begin{equation}
    \mu_{\text{X}}(\phi)= k_{\text{B}}T\bigg(\int\frac{Z_{\text{X}}(\phi)}{\phi}\,\mathrm{d}\phi + Z_{\text{X}}(\phi) -V_{\text{c}}n_{\text{R}}\diff{\alpha}{\phi}\bigg)+\text{const},\label{mucC_def}
\end{equation}
where $V_{\text{c}}=4\pi a^3/3$ is the volume of a colloid particle. The expression for the hard-sphere compressibility $Z_{\text{F}}(\phi)$ for the colloid-fluid phase is taken from Carnahan \& Starling~\cite{Carnahan1970}, and $Z_{\text{S}}(\phi)$ for colloid-crystal (hcp) phase is taken from the molecular dynamics simulations of Speedy~\cite{Speedy1998}:
\begin{align}
    Z_{\text{F}}(\phi)&=\frac{1+\phi+\phi^2-\phi^3}{(1-\phi)^3},\nonumber \\
    Z_{\text{S}}(\phi)&=\frac{3}{1-\phi/\phi_{\text{max}}} - 0.5935\Bigg(\frac{\phi/\phi_{\text{max}}-0.7080}{\phi/\phi_{\text{max}}-0.601}\Bigg),
\end{align}
where $\phi_{\text{max}}=\sqrt{2}\pi/6$ is the maximum packing fraction. We note that, while experiments have shown that the colloidal crystals in the fcc phase are more prevalent after a long time~\cite{Pusey1989} and in a microgravity environment~\cite{Cheng2001,Lei2024}, simulations have shown the entropies of the fcc and hcp phases to be very similar~\cite{Woodcock1997,Bolhuis1997,Mau1999}. We thus expect the difference between the two phases to have a negligible influence on results presented herein.

The constant term in Eq.~\eqref{mucC_def} is chosen to equate the fluid and solid chemical potentials at the coexistence compositions in the absence of polymer, $\mu_{\text{F}}(\phi_1)=\mu_{\text{S}}(\phi_2)$. Finally, the function $\alpha(\phi)$ in Eq.~\eqref{mucC_def} represents the free volume of the polymer coils~\cite{Lekkerkerker1992}, given by 
\begin{align}
&\alpha(\phi)= (1-\phi)\exp\left[-A\gamma-B\gamma^2-C\gamma^3\right],\,\text{where}\,\gamma=\frac{\phi}{1-\phi},\nonumber \\
&A = \xi(\xi^2+3\xi+3),\quad B = \frac{9\xi^2}{2}+3\xi^3,\quad C = 3\xi^3.
\end{align}
The parameter $n_{\text{R}}$ in Eq.~\eqref{mucC_def} is the ratio of the polymer number density $\rho N_{\text{A}}/M_{\text{p}}$ to the free volume $\alpha(\phi)$, where $N_{\text{A}}$ is Avogadro's number and $M_{\text{p}}=13.2\times 10^6$ g/mol is the polymer's molecular weight. The polymer chemical potential, and thus the parameter $n_{\text{R}}$, is assumed constant because the polymer diffuses quickly relative to the colloid~\cite{Lekkerkerker1992}.

In the Stokes equations~\eqref{Stokes}, $\bs{u}=\bs{u}(\bs{x},t)$ is the fluid velocity, $p=p(\bs{x},t)$ the pressure and $\eta(\phi)$ the dynamic viscosity. We note that the flow Reynolds number $\text{Re}=\rho U\ell/\eta=O(10^{-5})$ using the approximate values $\rho\sim 1\text{ g/cm}^3$ for the fluid density, $\eta\sim 10^{-2}\text{ g/(cm$\cdot$s)}$ for the solvent's dynamic viscosity, $\ell\sim 0.1$ cm for the characteristic domain size and $U\sim 10^{-6}$ cm/s for the flow speed, the last two of which are estimated from Fig.~\ref{fig:BCAT5Auto}. The influence of fluid inertia is thus neglected in Eq.~\eqref{Stokes}. Our model accounts for the fact that the viscosity of a colloidal suspension increases dramatically with 
$\phi$~\cite{Cheng2002}, in that we adopt the expression~\cite{Hunter2012}
\begin{align}
\frac{\eta(\phi)}{\eta_0}=\exp\left(\frac{D_{\text{v}}\phi}{\phi_{\text{v}}-\phi}\right)\label{HuntWeeksEta}
\end{align}
which, for $D_{\text{v}}=1.15$ and $\phi_{\text{v}}=0.638$, exhibits reasonably good agreement with the low shear viscosities of colloidal suspensions measured in experiments. The right-hand side of the Stokes equations contains the so-called Korteweg stress, which describes the stress exerted by the colloid on the fluid in a phase-field model~\cite{Jacqmin1999}. It has been demonstrated that the form of the Korteweg stress is consistent with measurements of the effective surface tension in colloidal suspensions that undergo a fingering instability~\cite{Truzolillo2014}. For a homogeneous binary fluid in which each component species has the same molar mass, we would have $\kappa = 1$~\cite{Lamorgese2017}; however, in this work the solvent (typically cis-decalin or tetralin) and solute (colloid particles) have different molar masses, making it difficult to obtain a simple expression for the stress~\cite{Lamorgese2011}. We thus take $\kappa$ as a free parameter in our model.

We non-dimensionalize the governing equations~\eqref{CHStokes} according to $\bs{x}\rightarrow\bs{x}/L$, $t\rightarrow t/\tau$, $\bs{u}\rightarrow\bs{u}/U$ and $p\rightarrow pL/(\eta_0U)$. The length, time and velocity scales are
\begin{align}
L = \frac{\ell}{\sqrt{\nu}},\quad \tau = \frac{\ell^2}{\nu^2\Gamma_0k_{\text{B}}T}\quad\text{and}\quad U = \frac{\kappa k_{\text{B}}T\ell^2}{\eta_0V_{\text{c}}L},\label{NDimScales}
\end{align}
respectively, where,
\begin{align}
\nu = \frac{|\mu^{\prime}(\phi_{\text{s}})|}{k_{\text{B}}T}\quad\text{and}\quad\phi_{\text{s}}=\text{argmax}\{|\mu^{\prime}(\phi)|:\mu^{\prime}(\phi) < 0\}
\end{align}
is the colloid concentration within the spinodal region for which the derivative of the chemical potential is most negative. Defining the Peclet number $\gamma$ as
\begin{align}
\gamma = \frac{U\tau}{L}=\frac{9\kappa}{2\nu}\left(\frac{\ell}{a}\right)^2,\label{Peclet}
\end{align}
and rescaling the mobility, chemical potential and suspension viscosity as 
\begin{align}
\Gamma\rightarrow\frac{\Gamma}{\Gamma_0},\quad\mu\rightarrow \frac{\mu}{|\mu^{\prime}(\phi_{\text{s}})|}\quad\text{and}\quad \eta\rightarrow \frac{\eta}{\eta_0},
\end{align}
we obtain the dimensionless equations
\begin{subequations}\label{CHStokesND}
\begin{align}
&\left(\partial_t+\gamma\bs{u}\cdot\bs{\nabla}\right)\phi=\bs{\nabla}\cdot\left[\Gamma(\phi)\bs{\nabla}\left(\mu(\phi)-\Delta\phi\right)\right],\label{CHND} \\
&\bs{\nabla}\cdot\Big[\eta(\phi)\Big(\nabla\bs{u} + (\nabla\bs{u})^{\text{T}}\Big)\Big]=\bs{\nabla}p+\bs{\nabla}\cdot\left(\bs{\nabla}\phi\bs{\nabla}\phi\right),\label{StokesND} \\
 &\bs{\nabla}\cdot\bs{u}=0.\label{IncompND}
\end{align}
\end{subequations}
The dimensionless equations thus depend on the single free parameter $\gamma$, as defined in Eq.~\eqref{Peclet}, which describes the relative importance of advective and diffusive transport of the colloids.

\section{Numerical implementation}\label{Sec:Numerics}

We solve the governing equations~\eqref{CHStokesND} numerically with doubly periodic boundary conditions on a square domain of size $4\pi P\times 4\pi P$, where $P=20$ unless otherwise stated. A linear stability analysis of Eqs.~\eqref{CHStokesND} around the homogeneous state $\phi=\phi_0$ in the spinodal region and $\bs{u}=0$ yields the upper bound $|\bs{k}| = 1/\sqrt{2}$ for the wavenumber of maximum growth, which corresponds to a wavevector $\bs{k}=(k_x,k_y)$ with components $k_x=k_y=1/2$, for example. Our choice of simulation domain thus admits $P$ such wavelengths along each direction. To mimic the initially homogenized state of the suspension immediately after mixing, we generate a small-amplitude field of Gaussian random perturbations $\tilde{\phi}(\bs{x})$ with zero mean, smooth it, and add to it the homogeneous volume fraction $\phi_0$, thus obtaining the initial condition $\phi(\bs{x},0)=\phi_0+\tilde{\phi}(\bs{x})$. The parameter $n_{\text{R}}$ that determines the polymer chemical potential is taken to be that corresponding to the average colloid volume fraction:
\begin{align}
n_{\text{R}}=\frac{\rho N_{\text{A}}M_{\text{p}}}{\alpha(\phi_0)}.
\end{align}
When $\phi_0$ is in the spinodal region, $\mu^{\prime}(\phi_0) < 0$, these small-amplitude perturbations will grow in time and spinodal decomposition will take place.

We employ a pseudospectral method with 512 points in each direction. We proceed by describing, in turn, the time-stepping scheme for solving the Cahn-Hilliard equation~\eqref{CHND} (\S\ref{ssec:time}) and the method for solving the variable-viscosity Stokes equations~\eqref{StokesND}-\eqref{IncompND} (\S\ref{ssec:Stokes}).

\subsection{Semi-implicit time-stepping scheme}\label{ssec:time}

The first-order forward Euler time-stepping method for Eq.~\eqref{CHND}, written in Fourier space, is
\begin{align}
    &\hat{\phi}^{n+1} = \hat{\phi}^n - \gamma\Delta t\mathcal{F}\left[\bs{u}^n\cdot\bs{\nabla}\phi^n\right]\nonumber \\
    &+ \Delta t \,\rmi\bs{k}\cdot\mathcal{F}\left[ \Gamma(\phi^n)\mathcal{F}^{-1}\left[ \rmi\bs{k}\left( \mathcal{F}\left[\mu(\phi^n)\right] + |\bs{k}|^2\hat{\phi}^n \right) \right] \right],\label{CHForwardEuler}
\end{align}
where the hats are used to denote Fourier-transformed variables, $\hat{\phi}=\mathcal{F}[\phi]$, $\hat{\phi}^n=\hat{\phi}(\bs{k},n\Delta t)$ and $\Delta t$ is the time step. Equation~\eqref{CHForwardEuler} suffers from a severe time-stepping restriction, $\Delta t \sim O((\Delta x)^4)$. To avoid this problem, Zhu {\it et al.}~\cite{Zhu1999} proposed a semi-implicit time-stepping scheme. Specifically, they add and subtract a constant $A$ from the mobility,
\begin{equation}
    \Gamma(\phi)=\big(\Gamma(\phi)-A\big)+A,
\end{equation}
which results in a linear term $-A\Delta^2\phi$ on the right-hand side of Eq.~\eqref{CHND}. This numerically stiff term is treated implicitly, while all of the other (nonlinear) terms in the equation are treated explicitly as in Eq.~\eqref{CHForwardEuler}. While Zhu {\it et al.} suggested $A=\big( \max(\Gamma(\phi)) + \min(\Gamma(\phi)) \big)/2$, we found by experimenting with the numerical scheme that $A=1$ is a suitable choice for the mobility defined in Eq.~\eqref{GammaDef}.
Rearranging the resulting equation, we obtain the time-stepping scheme
\begin{align}
    &\hat{\phi}^{n+1} = \hat{\phi}^n +\frac{\Delta t}{1+A\Delta t |\bs{k}|^4}\Big(  -\gamma\mathcal{F}\left[ \bs{u}^{n}\cdot\bs{\nabla}\phi^{n} \right]\nonumber\\
    &+ \rmi\bs{k}\cdot\mathcal{F}\left[ \Gamma(\phi^{n})\mathcal{F}^{-1}\left[\rmi\bs{k}\left(\mathcal{F}\left[\mu(\phi^{n})\right]+|\bs{k}|^2\hat{\phi}^{n} \right) \right]  \right]   \Big).\label{Zhu1stOrd}
\end{align}

\subsection{Solving the variable-viscosity Stokes equations}\label{ssec:Stokes}

We adapt the procedure of Tree {\it et al.}~\cite{Tree2017} to solve the variable-viscosity Stokes equations~\eqref{StokesND}-\eqref{IncompND} at every time step. Dropping the superscripts on $\phi$ for the sake of clarity, we write
\begin{equation}
    \eta(\phi)=\eta^* + \bar{\eta}(\phi), \quad\text{where}\quad \eta^*=\max\eta(\phi),
\end{equation}
and define the tensors
\begin{align}
\bs{\tau}=\bs{\nabla}\bs{u}+(\bs{\nabla}\bs{u})^{\text{T}}\quad\text{and}\quad\bs{\sigma}=\bs{\nabla}\phi\bs{\nabla}\phi
\end{align}
to rewrite Eq.~\eqref{StokesND} in the form
\begin{align}
\eta^*\Delta\bs{u} &= \bs{\nabla} p - \bs{\nabla}\cdot\big[\bar{\eta}\boldsymbol{\tau} \big]+\nabla\cdot\boldsymbol{\sigma}.\label{Stokes1}
\end{align}
Taking the Fourier transform of Eqs.~\eqref{Stokes1} and using the incompressibility condition, Eq.~\eqref{IncompND}, we obtain~\cite{Tree2017}
\begin{align}
 \hat{\bs{u}} = -\frac{1}{\eta^*|\bs{k}|^2}\left(\bs{I}-\frac{\bs{k}\bs{k}^{\text{T}}}{|\bs{k}|^2} \right)    \left[\rmi\bs{k}\cdot\left( -\mathcal{F}\left[\bar{\eta}\boldsymbol{\tau}\right] + \mathcal{F}\left[\boldsymbol{\sigma}\right] \right)\right],\label{Eq86}
\end{align}
where $\bs{I}$ is the identity matrix. Defining the scalar variable $\hat{u}$ through the equation
\begin{align}
\hat{\bs{u}}=\frac{\rmi\bs{k}^\perp}{|\bs{k}|}\hat{u}\quad\text{where}\quad \bs{k}^{\perp}=(k_y,-k_x),\label{uHatDef}
\end{align}
and using the fact that $\bs{I}-\bs{k}\bs{k}^{\text{T}}/|\bs{k}|^2=\bs{k}^{\perp}\bs{k}^{\perp\text{T}}/|\bs{k}|^2$, we obtain a linear equation for $\hat{u}$:
\begin{align}
\hat{u} = \frac{\rmi}{\eta^*|\bs{k}|^3}\bs{k}^\perp\cdot\left[ \rmi\bs{k}\cdot\left( -\mathcal{F}\left[\overline{\eta}\boldsymbol{\tau}\right] + \mathcal{F}\left[\boldsymbol{\sigma}\right] \right) \right]. \label{uHatEq1}
\end{align}
Our approach differs from that of Ref.~\cite{Tree2017} in that the scalar Eq.~\eqref{uHatEq1} is solved, rather than the vector Eq.~\eqref{Eq86}, which reduces the problem's computational cost. 

Note that Eq.~\eqref{uHatEq1} can be written as
\begin{align}
u = H(u),\label{FixedPointProb}
\end{align}
where $u = \mathcal{F}^{-1}[\hat{u}]$ and $H$ is a linear operator. It is not feasible to solve Eq.~\eqref{FixedPointProb} explicitly using Gaussian elimination, for example, due to the fact that $u$ contains $512^2$ unknowns. For this reason, we solve the linear system using a combination of Picard iteration~\cite{Lamorgese2017} and Anderson mixing~\cite{Ng1974,Thompson2004,Tree2017}, iterative methods that allow us to approximate the solution to Eq.~\eqref{FixedPointProb} with a desired accuracy. We briefly describe how to get from step $k$ to $k+1$ of the iteration: suppose for now that $k\geq J$, a predetermined number, and that we have the $J+1$ approximations (or ``iterates'') $u_{k-J},\dots,u_k$. The Anderson mixing scheme reads
\begin{align}
u_{k+1}&=H(\bar{u}_{k+1})\quad\text{for}\quad k\geq J,\nonumber \\
\text{where}\quad \bar{u}_{k+1} &= u_k + \sum_{j=1}^J c_j^{(k)}\big( u_{k-j} - u_k\big).\label{AndMix}
\end{align}
The constants $\bs{c}^{(k)}=\left(c_1^{(k)},\dots,c_J^{(k)}\right)$ are obtained by solving the linear system
\begin{align}
U^{(k)}\bs{c}^{(k)}=\bs{v}^{(k)},\label{LinSystAndMix}
\end{align}
where $U^{(k)}\in\mathbb{R}^{J\times J}$ and $\bs{v}^{(k)}\in\mathbb{R}^J$ have the elements
\begin{align}
U^{(k)}_{ij}&=\left\langle d_k^{(k)}-d_{k-i}^{(k)}, d_k^{(k)}-d_{k-j}^{(k)}\right\rangle,\nonumber \\
v^{(k)}_i &= \left\langle d_k^{(k)}-d_{k-i}^{(k)}, d_k^{(k)}\right\rangle,\quad 1\leq i,j\leq J,
\end{align}
and
\begin{align}
d_{l}^{(k)}=H(u_{l})-u_{l},\quad k-J\leq l\leq k,
\end{align}
$\langle f,g\rangle=\int f(\bs{x})g(\bs{x})\,\rmd\bs{x}$ being the standard inner product. The iteration, Eq.~\eqref{AndMix}, continues until the error residual $\left\|d_k^{(k)}\right\|\equiv \sqrt{\left\langle d_k^{(k)},d_k^{(k)}\right\rangle}$ drops below the error tolerance $10^{-6}$.

We proceed by describing how we obtain the first $J+1$ iterates $u_0,u_1,\dots,u_J$. Starting with $u_0$, for the first time step of the simulation, we use the solution to the {\it constant}-viscosity Stokes equations, which have the analytical solution in Fourier space
\begin{align}
\hat{u}_0=\frac{1}{\eta(\phi_0)|\bs{k}|^3}\rmi\bs{k}^\perp\cdot\left(\rmi\bs{k}\cdot\mathcal{F}[\boldsymbol{\sigma}]\right),\label{StokesConstEta2}
\end{align}
and let $u_0 = \mathcal{F}^{-1}[\hat{u}_0]$. For every time step after the first, we let $u_0$ be the solution to the problem Eq.~\eqref{FixedPointProb} at the previous timestep. Given $u_0$, we obtain the $J$ iterates $u_1,\dots,u_J$ using Picard iteration with relaxation:
\begin{align}
u_{k+1}=\omega H(u_k) + (1-\omega)u_k, \quad 0\leq k < J,\label{Picard}
\end{align}
where we choose the relaxation parameter to be $\omega=0.1$. This completely defines an iterative scheme for solving Eq.~\eqref{FixedPointProb}.

While the iterative scheme, Eq.~\eqref{AndMix}, typically converges faster for larger values of $J$, the computational cost of performing a single iteration clearly increases with $J$. To handle this tradeoff in our numerical simulations, we found that $J=2$ was sufficient for early times $t$, when the constant viscosity solution $u_0$ is a good approximation to the actual solution to Eq.~\eqref{FixedPointProb}. As coarsening progresses and $\eta$ is no longer nearly constant, our algorithm automatically increases $J$ if convergence is not reached within $40+10(J-2)$ iterations, where we count both Picard iterations, Eq.~\eqref{Picard}, and Anderson mixing ones, Eq.~\eqref{AndMix}. 

\section{Results of the numerical simulations}\label{Sec:results}

\begin{figure*}[hbtp]
    \centering
    \includegraphics[width=0.75\textwidth]{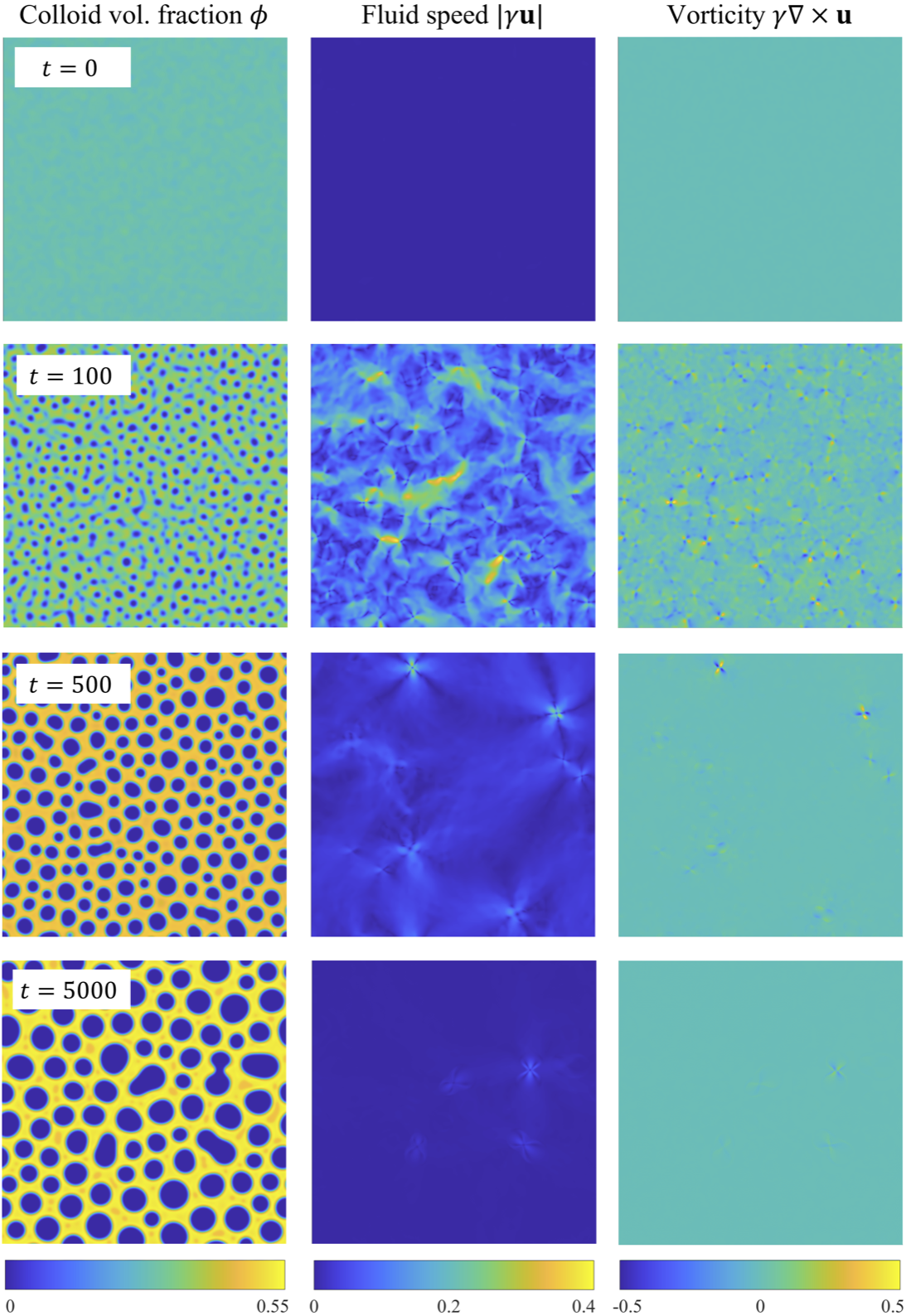}
    \caption{Simulation results for BCAT-5 Sample 7, with $\gamma=130$. 
    Each row represents a different (dimensionless) time $t$ as labeled, with the three panels in each row corresponding to colloid volume fraction $\phi$, fluid speed $|\gamma \bs{u}|$, and fluid vorticity $\gamma\bs{\nabla}\times\bs{u}$. In dimensional variables, the simulation domain length is roughly 10 mm and the final time of the simulation 295 hours. The maximum fluid speed in the color bar is roughly 0.1 $\mu$m/s, and the maximum vorticity $2\times 10^{-3}$ s$^{-1}$. A video of this simulation is available in Supplementary Movie 1.}
    \label{fig:SimSnaps_Samp7}
\end{figure*}

We proceed by showing the results of numerical simulations of the phase-field model, Eq.~\eqref{CHStokesND}, as conducted using the procedures described in \S\ref{Sec:Numerics}. The parameters correspond to the BCAT-5 experiments, as given in Table~\ref{tab:param}. Supplementary Movie 1 shows a simulation with parameters corresponding to BCAT-5 Sample 7, with hydrodynamics included ($\gamma=130$). Snapshots of the simulation at the four (dimensionless) times $t=0$, 100, 500 and 5000 are shown in Fig.~\ref{fig:SimSnaps_Samp7}, with the left column showing the colloid volume fraction $\phi(\bs{x},t)$. 

\begin{figure}[hbtp]
    \centering
    \includegraphics[width=1\columnwidth]{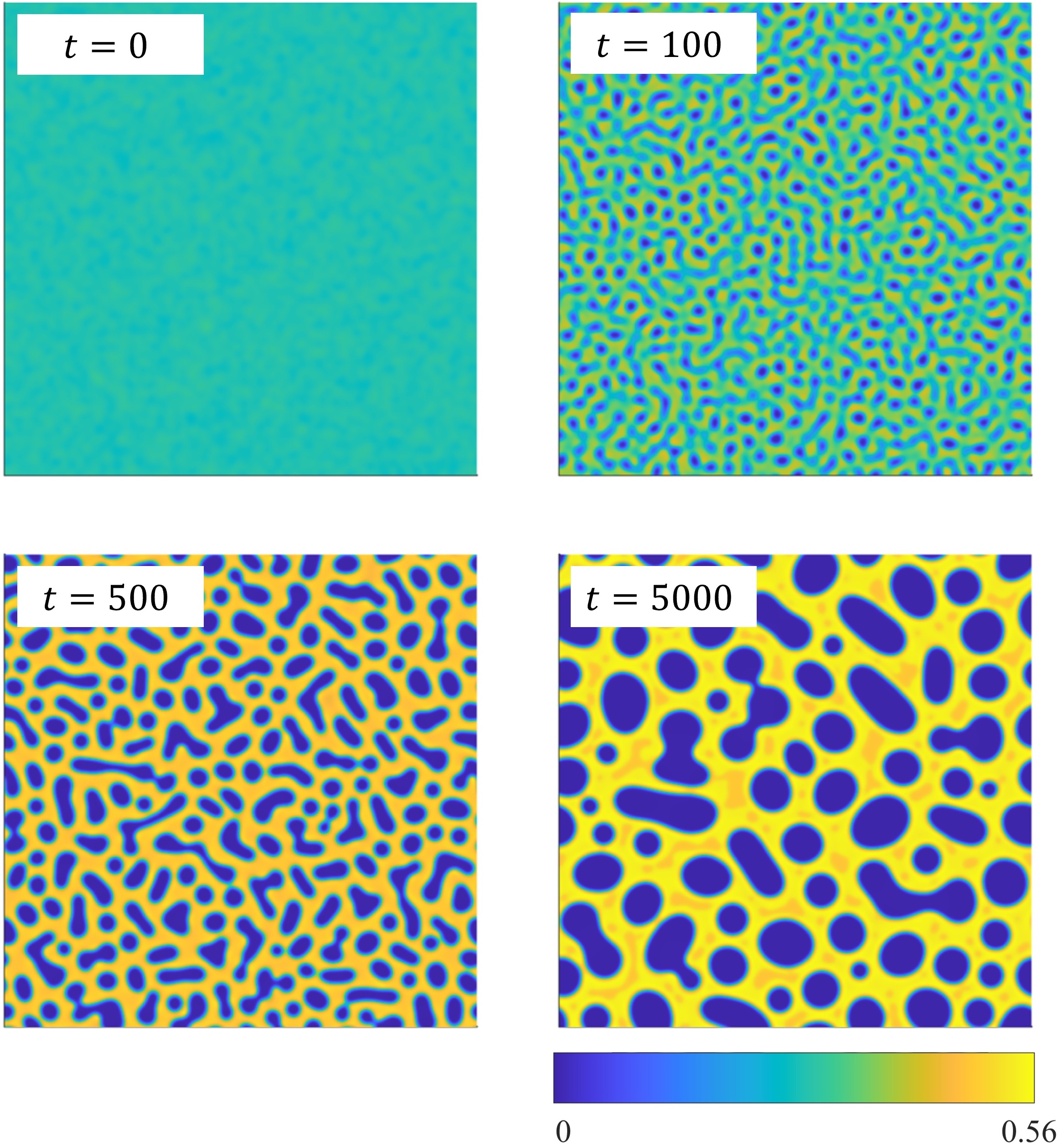}
    \caption{Simulation results for BCAT-5 Sample 7, without hydrodynamic effects ($\gamma=0$). Each panel represents the colloid volume fraction $\phi$ at a different (dimensionless) time $t$, as labeled. In dimensional variables, the simulation domain length is roughly 10 mm and the final time of the simulation 295 hours. The bottom left panel of Supplementary Movie 3 shows a video of this simulation.}
    \label{fig:SimSnaps_Samp7NOFLUID}
\end{figure}

At $t=0$, the mixture is nearly homogeneous, $\phi\approx\phi_0=0.29$, with only small perturbations present. By $t=100$, small domains of low-$\phi$ ``gas'' phase (dark blue) have formed; higher-$\phi$ (lighter green) regions also become visible. The phase domains continue to grow, and the higher-$\phi$ ``liquid'' (orange) values take over, forming the background as the low-$\phi$ ``droplets'' merge to form larger gas domains. This liquid state, however, is temporary, as it is only a metastable state. Eventually, crystallization sets in, and there is a three-phase coexistence of solid, liquid, and gas. By $t=5000$, the solid phase has mostly taken over, though small regions of the liquid phase remain. The gas phase ``droplets'' continue to merge, and coarsening progresses at a relatively slow pace. The simulation thus exhibits a three-phase coexistence of solid, gas, and a metastable liquid phase, which has been predicted theoretically in prior work~\cite{Evans2001} and observed experimentally~\cite{Ilett1995,Poon1999}. As expected, three-phase coexistence is observed even if hydrodynamic effects are neglected ($\gamma=0$), as shown in Fig.~\ref{fig:SimSnaps_Samp7NOFLUID}. 

\begin{figure*}[hbtp]
    \centering
    \includegraphics[width=.75\textwidth]{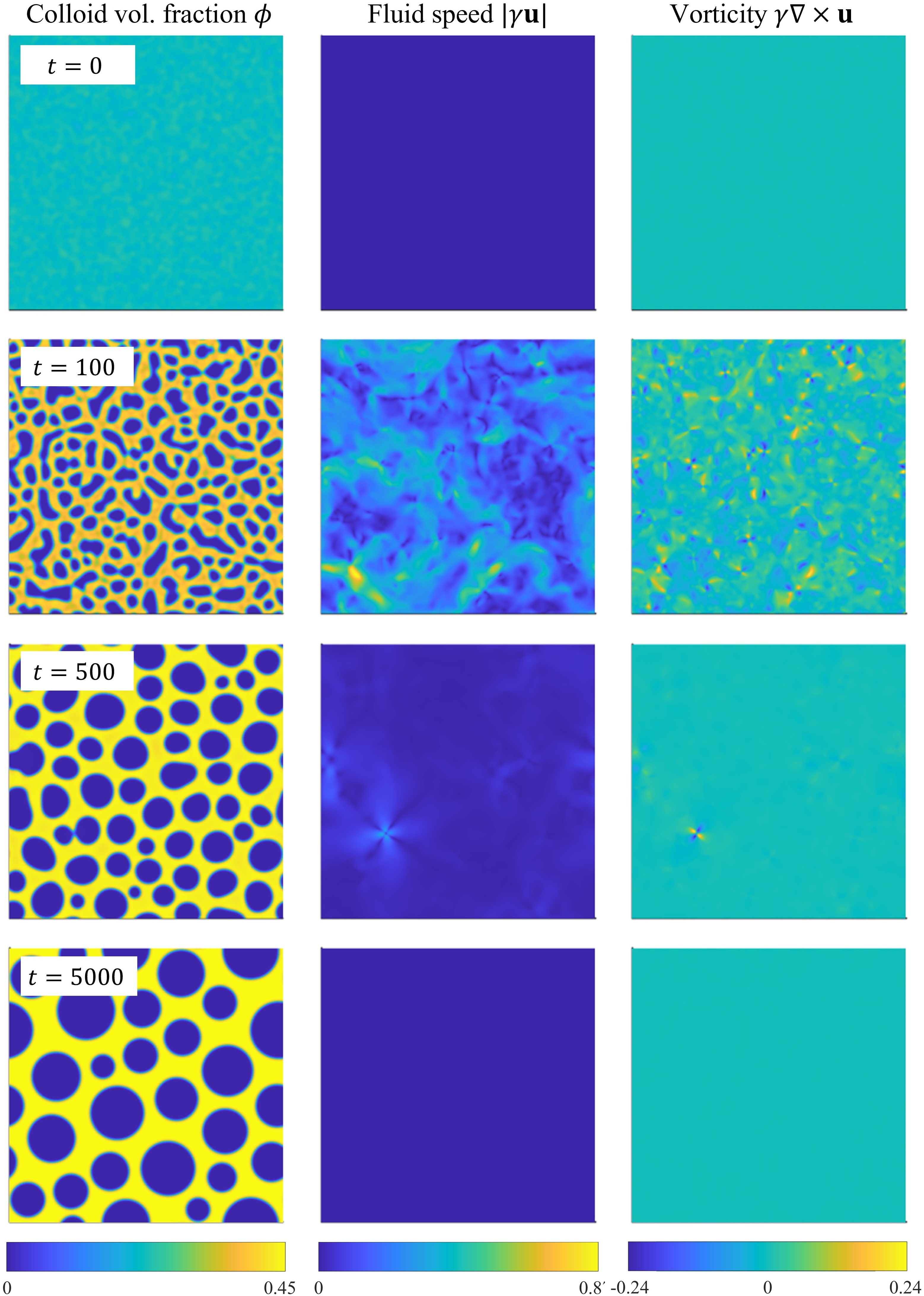}
    \caption{Simulation results for BCAT-5 Sample 4, with $\gamma=93$. 
    In dimensional variables, the simulation domain length is roughly 14 mm and the final time of the simulation 1100 hours. The maximum fluid speed in the color bar is roughly 0.06 $\mu$m/s, and the maximum vorticity $3\times 10^{-4}$ s$^{-1}$. A video of this simulation is available in Supplementary Movie 2.}
    \label{fig:SimSnaps_Samp4}
\end{figure*}

\begin{figure}[hbtp]
    \centering
    \includegraphics[width=1\columnwidth]{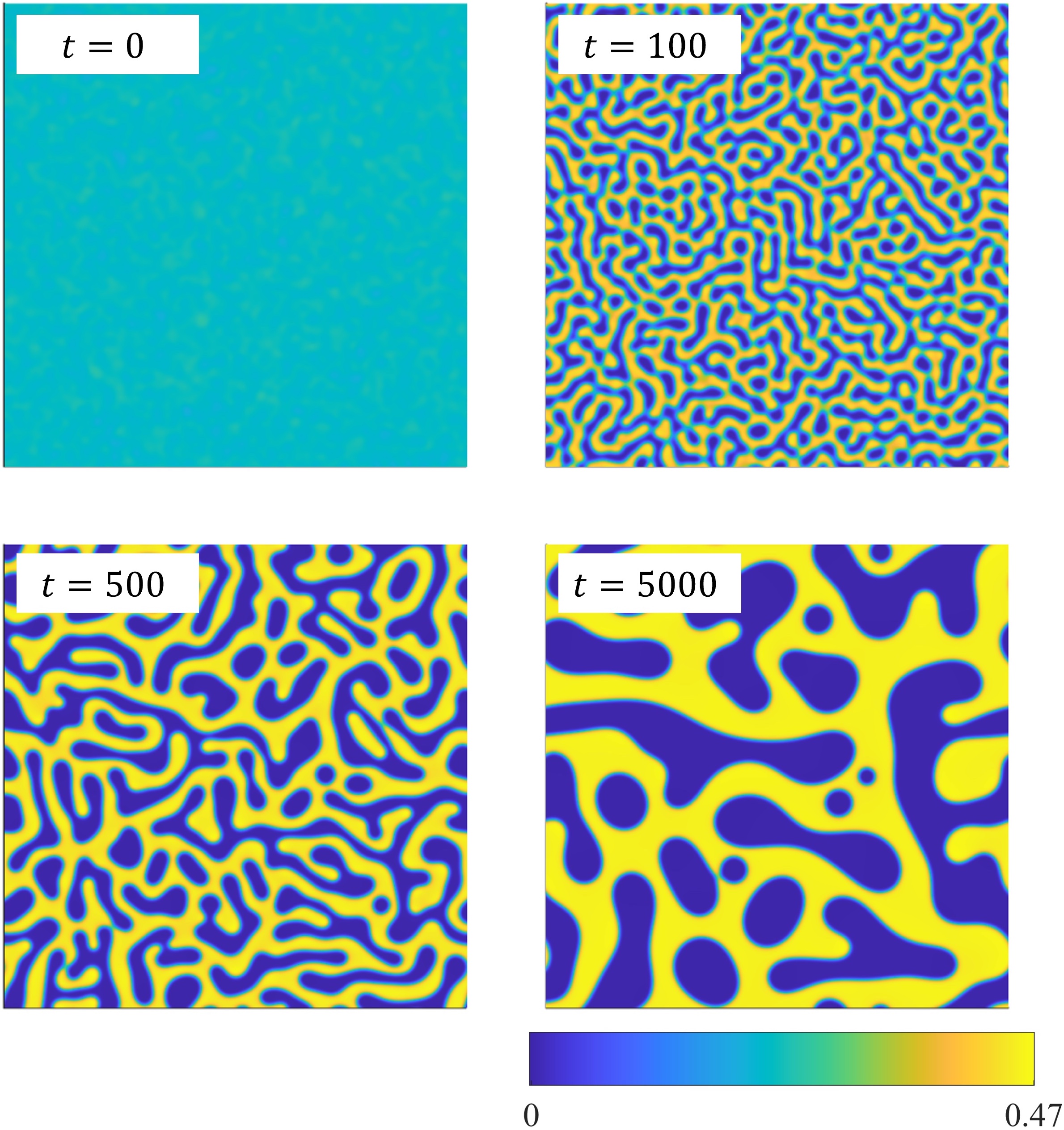}
    \caption{Simulation results for BCAT-5 Sample 4, without hydrodynamic effects ($\gamma=0$). In dimensional variables, the simulation domain length is roughly 14 mm and the final time of the simulation 1100 hours. The top left panel of Supplementary Movie 3 shows a video of this simulation.}
    \label{fig:SimSnaps_Samp4NOFLUID}
\end{figure}

We observe that including hydrodynamic interactions between the colloid particles and the surrounding fluid in the model seems to make the phase domains rounder in shape. This effect is most visible in the simulation of BCAT-5 Sample 4, which is shown in Fig.~\ref{fig:SimSnaps_Samp4} and Supplementary Movie 2 for $\gamma=93$. Especially at late times, the phase domains shown in the left column of Fig.~\ref{fig:SimSnaps_Samp4} appear more circular than those in Fig.~\ref{fig:SimSnaps_Samp4NOFLUID}, for which hydrodynamic effects were not incorporated ($\gamma=0$). This phenomenon is illustrated in Supplementary Movie 3, where the time-evolution of BCAT-5 Samples 4 and 7 with and without hydrodynamics are compared side-by-side. Our finding is somewhat at odds with the results of Ref.~\cite{Tanaka2000a}, who conducted simulations of colloids interacting through a short-range Lennard-Jones potential by modeling colloids as fluid regions of relatively large viscosity. They found that including hydrodynamic interactions resulted in the formation of chainlike aggregates instead of circular clusters. We attribute the discrepancy with our results to differences in the modeling framework; specifically, the Cahn-Hilliard equation is not solved in Ref.~\cite{Tanaka2000a} so spinodal decomposition is not modeled explicitly, as it is in Eq.~\eqref{CH}.

The time evolution of the (dimensionless) fluid speed $|\gamma\bs{u}|$ is shown in the middle column of Figs.~\ref{fig:SimSnaps_Samp7} and~\ref{fig:SimSnaps_Samp4}, and Supplementary Movies 1 and 2. 
The initial gradients in $\phi$ are small, corresponding to low fluid speeds. The fluid speeds increase quickly thereafter, and are largest for early times (i.e. $t=100$), slowing down at later times (i.e. $t=500$ and $1000$). 

\begin{figure}[hbtp]
    \centering
    \includegraphics[width=1\columnwidth]{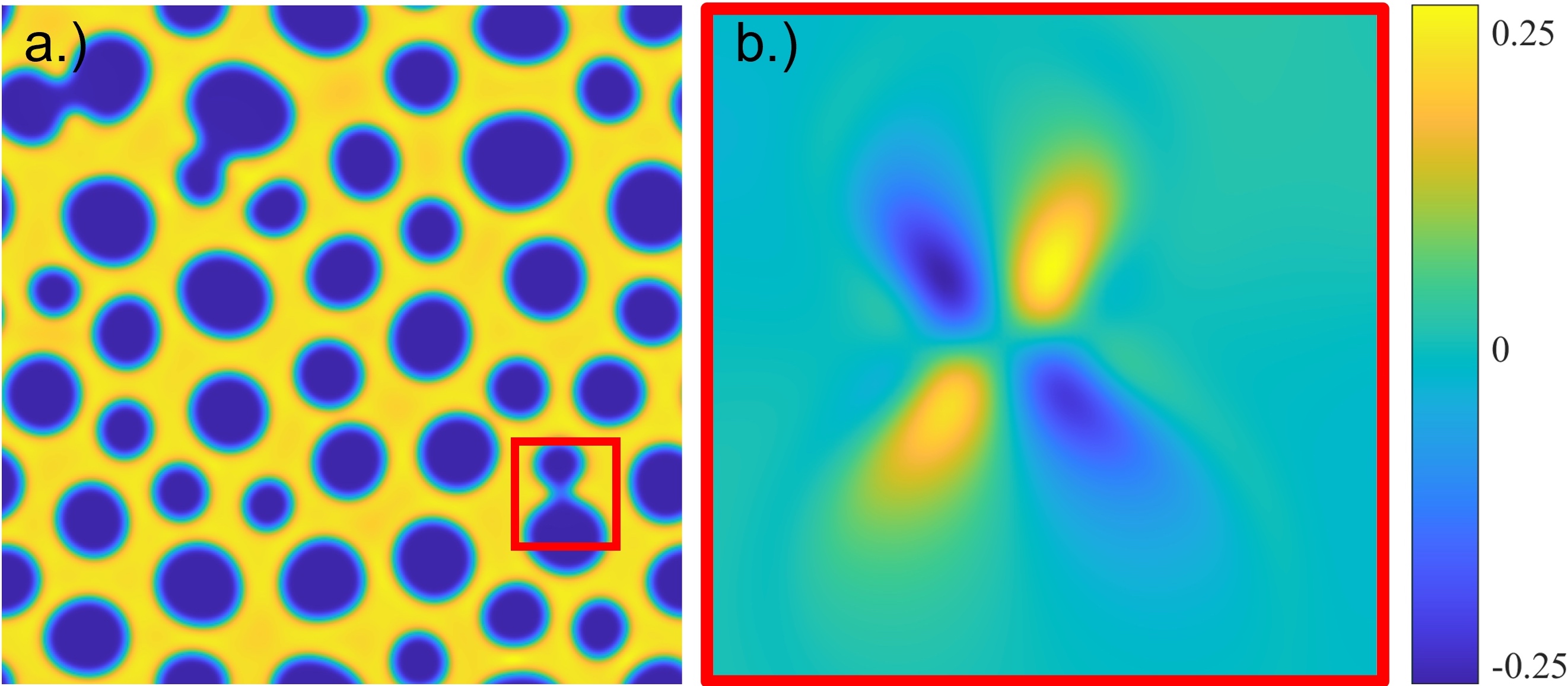}
    \caption{(a) Plot of $\phi$ for a BCAT-5 Sample 7 simulation using $\gamma=130$ 
    and $P=10$ at (dimensionless) time $t=527$. (b) Zoom-in of the fluid vorticity $\gamma\bs{\nabla}\times\bs{u}$ in the region enclosed by the red box in panel (a).}
    \label{fig:VortQuad}
\end{figure}

Another phenomenon we observed in our simulations with hydrodynamics ($\gamma>0$) is the formation of a quadrupole of vortices whenever two phase domains merge, as illustrated in Fig.~\ref{fig:VortQuad}.  Each vortex quadrupole consists of two pairs of vortices: one pair is clockwise (blue) and the other counter-clockwise (yellow), positioned across from each other. Such structures are also visible in Figs.~\ref{fig:SimSnaps_Samp7} and~\ref{fig:SimSnaps_Samp4}, specifically, at $t=100$ when there is a merger between colloid-poor domains. They are also visible in the right columns of Supplementary Movies 1 and 2, which show vortex quadrupoles popping in and out of existence as phase separation progresses. While the fluid is relatively quiescent at late times when the coarsening slows, the vortex quadrupoles continue to appear whenever there is a merger of domains. 

\section{Comparison between theory and experiment}\label{Sec:Theoryvsexp}

\begin{figure}[!htbp]
    \centering
    \includegraphics[width=1\columnwidth]{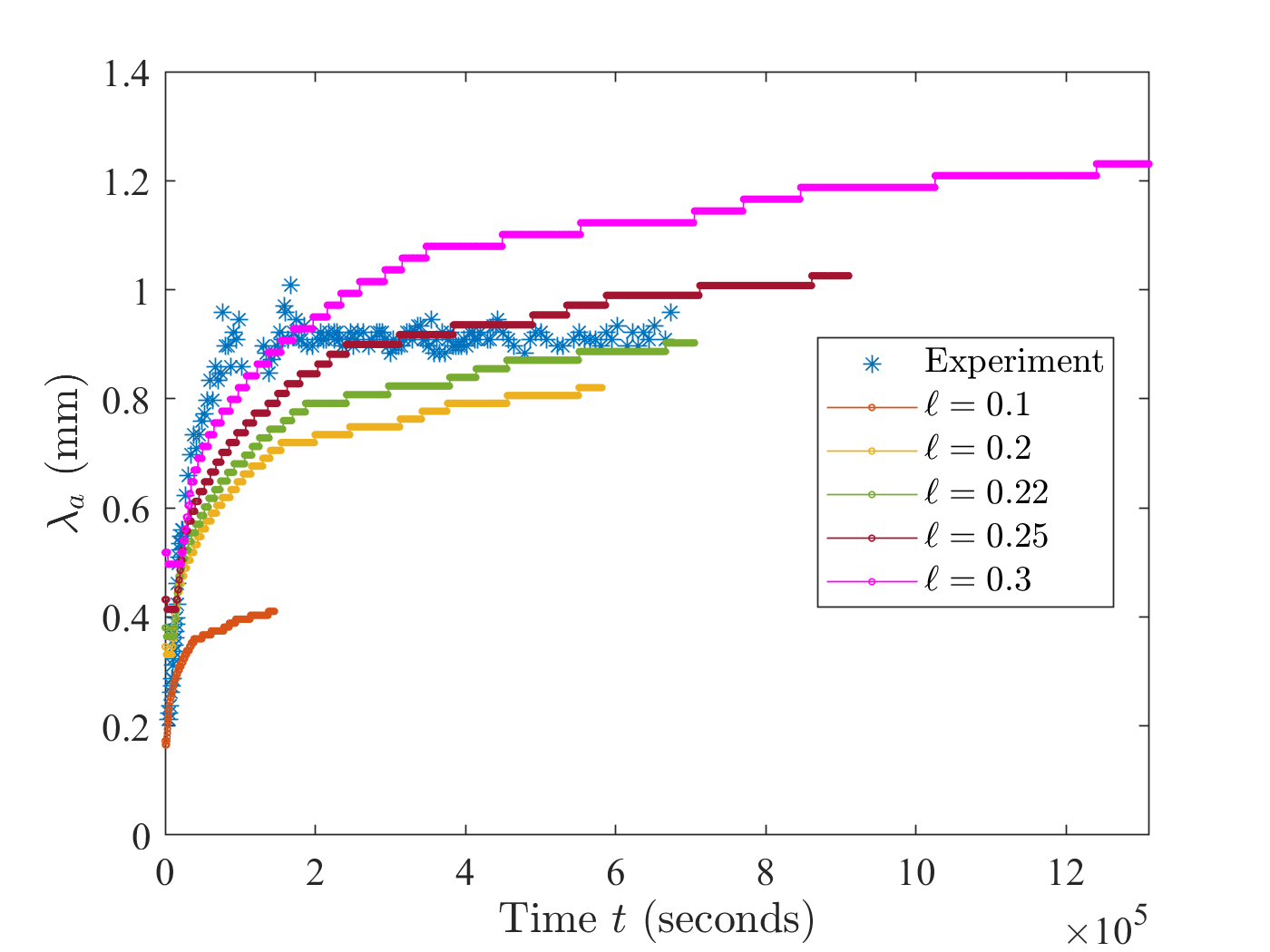}
    \caption{Dominant length scale $\lambda_a$ vs. $t$ for BCAT-5 Sample 7, obtained from numerical simulations using $\gamma=130$, 
    plotted with the experimental data (blue points). The different colors correspond to different choices of $\ell$ (in mm) used in computing the length and time scales $L$ and $\tau$ defined in Eq.~\eqref{NDimScales}.}
    \label{fig:AutocorBCAT5Samp7_ManyElls}
\end{figure}

After carrying out the simulations with parameters corresponding to the BCAT-5 samples, we proceeded to determine the time evolution of the characteristic length scale $\lambda_a(t)$ using the algorithm described in \S\ref{Sec:Image}, 
in order to obtain a quantitative description of the coarsening process. To determine an appropriate value for the free parameter $\ell$, we experimented by plugging in various values for $\ell$ and plotting the corresponding curves (now in dimensional units) for each $\ell$ on the same axes as the experimental data, as shown in Fig.~\ref{fig:AutocorBCAT5Samp7_ManyElls}. The dimensional length and time scales were obtained using Eq.~\eqref{NDimScales}. We did this for each of the five BCAT-5 samples, selecting the $\ell$-value that best matched the experimental data. Then, we chose an $\ell$-value about midway between the highest and lowest values across the five samples, from which we obtained $\ell = 0.27$ mm. 

\begin{figure*}[hbtp]
    \centering
    \includegraphics[width=1\textwidth]{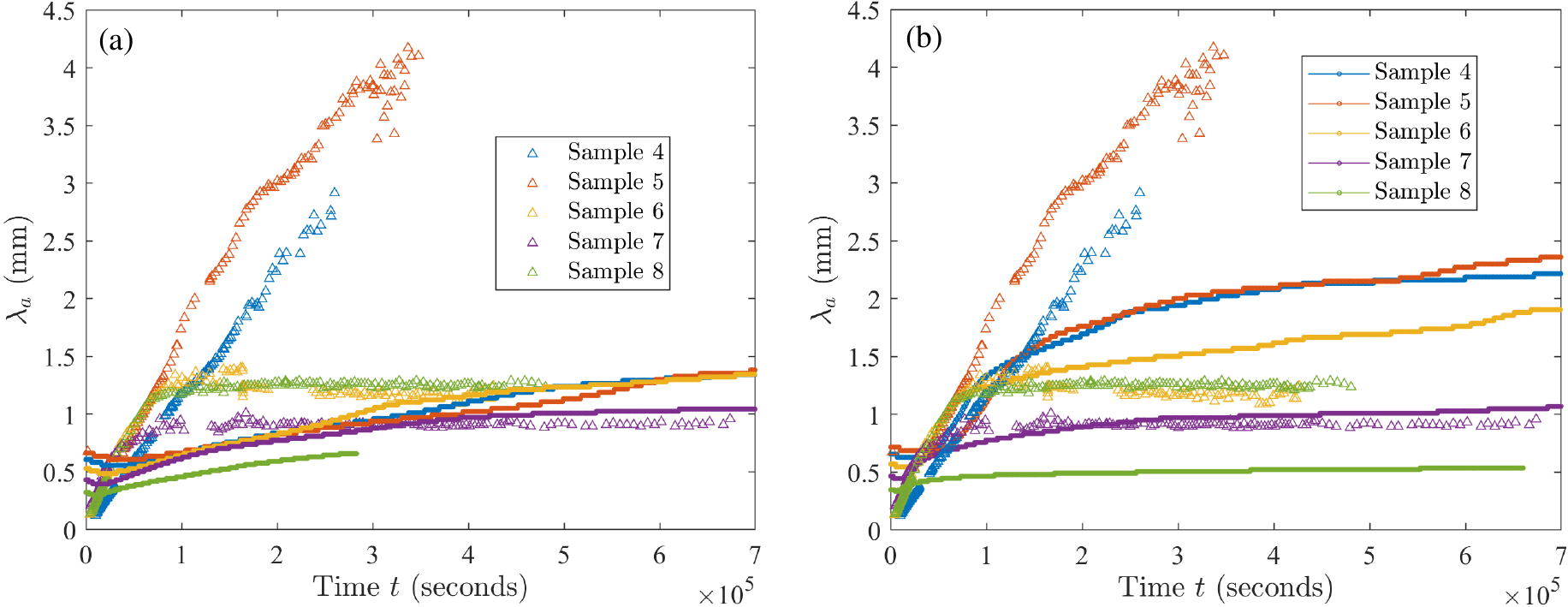}
    \caption{Dominant length scale $\lambda_a(t)$ in BCAT-5 Samples 4, 5, 6, 7 and 8, obtained from experimental data (triangles) and simulations  (solid curves) run without accounting for hydrodynamic interactions ($\gamma=0$, panel (a)) and with hydrodynamic interactions 
    (panel (b)). In the latter, the values of $\gamma$ are 93 (Sample 4), 91 (Sample 5), 100 (Sample 6), 130 (Sample 7) and 202 (Sample 8). The value $\ell=0.27$ mm is fixed.}\label{fig:AutocorrExp}
\end{figure*}

Fixing $\ell = 0.27$ mm, we proceeded to plot the growth curves for the BCAT-5 experiments in Fig.~\ref{fig:AutocorrExp}, comparing experimental data (triangles) with simulation data (solid curves) both without (Fig.~\ref{fig:AutocorrExp}(a)) and with (Fig.~\ref{fig:AutocorrExp}(b)) hydrodynamic interactions. We first consider the simulations with hydrodynamics (Fig.~\ref{fig:AutocorrExp}(b)). While the experimental and simulation data do not match perfectly, the values of $\lambda_a$ predicted by the simulations are the same order of magnitude as those in experiment. Moreover, the dependence of the long-time length scale on sample parameters is mostly consistent between experiment and simulation. Specifically, for the experimental data, the largest $\lambda_a$ values are attained by Sample 5 (red); then Sample 4 (blue); then Sample 6 (yellow) and Sample 8 (green), which are nearly the same; and then Sample 7 (purple). Sample 6 attains higher values overall than Sample 8, but its values eventually decrease, presumably due to artifacts arising from noise in the experimental images, and finally settle at a lower value than Sample 8. For the simulation data, the ordering is as follows: Sample 5 (red), Sample 4 (blue), Sample 6 (yellow), Sample 7 (purple), and finally Sample 8 (green), which is mostly consistent with the ordering of the experimental data. We note that we only had the freedom to adjust the single free parameter $\ell$ to obtain fits across all times $t$ and five BCAT-5 samples.

In the absence of hydrodynamic effects, $\gamma=0$ (Fig.~\ref{fig:AutocorrExp}(a)), the simulation curves for the five samples are clustered closer together than in Fig.~\ref{fig:AutocorrExp}(b); the values of $\lambda_a$ for Samples 4, 5 and 6 at the latest time considered are quite similar to each other,  
while Sample 7 attains significantly lower values and Sample 8 attains the lowest of all. Inclusion of hydrodynamic interactions appears to cause the $\lambda_a$-curves to separate from each other, making the different sample parameters more distinguishable from each other, as is the case in the experiments. 

We also note that the early-time growth rates of the simulation curves in Fig.~\ref{fig:AutocorrExp}(b) are more similar to the experimental data than those in Fig.~\ref{fig:AutocorrExp}(a), indicating that hydrodynamics affects the early-time behavior of the system.  Specifically, the $\gamma>0$ 
curve (Fig.~\ref{fig:AutocorrExp}(b)) for each sample exhibits faster growth than its $\gamma=0$ counterpart (Fig.~\ref{fig:AutocorrExp}(a)) at early times. Both curves begin to flatten out as time progresses, indicating a slowing down of the phase separation. This slowing starts earlier with hydrodynamics than without. These results suggest that the inclusion of hydrodynamic effects initially speeds up the phase separation process, but eventually slows it down at later times. Specifically, the increase in suspension viscosity, Eq.~\eqref{HuntWeeksEta}, and decrease in mobility, Eq.~\eqref{GammaFunc}, as the colloid volume fraction increases may lead to phase separation being hindered or even arrested in a colloid-polymer suspension. This mechanism would be in addition to the one identified by Sabin {\it et al.}~\cite{Sabin2012}, who analyzed microgravity experiments of colloid-polymer suspensions. They found that phase separation was arrested due to the emergence of a system-spanning crystal gel, an effect that cannot be captured by our model because it does not include gelation.

\section{Discussion}\label{Sec:Discussion}

We have constructed and analyzed a phase-field model, Eq.~\eqref{CHStokes}, for colloid-polymer mixtures in which hydrodynamic interactions are considered explicitly. The colloid volume fraction $\phi$ evolves according to the Cahn-Hilliard equation~\eqref{CH} with a concentration-dependent mobility $\Gamma(\phi)$, and the fluid velocity $\bs{u}$ is governed by the incompressible Stokes equations~\eqref{Stokes}-\eqref{Incomp} with a concentration-dependent viscosity $\eta(\phi)$. The fluid is forced by gradients in the colloid concentration via the Korteweg stress, which models the interfacial tension between colloidal phases. We have also processed video microscopy images from NASA's microgravity BCAT experiments, extracted the coarsening rates as a function of time, and quantitatively compared them with the predictions of our model.

Our simulation results provided some insight into the role of hydrodynamic effects on phase separation in colloid-polymer suspensions. Specifically, we observed that the phase domain structures take on a more stringy appearance when hydrodynamic interactions are neglected, as is evident in Supplementary Movie 3 and by comparing Figs.~\ref{fig:SimSnaps_Samp4} and~\ref{fig:SimSnaps_Samp4NOFLUID}. We also found that the merger of colloid-poor domains is associated with a vortex quadrupole in the fluid (Fig.~\ref{fig:VortQuad}). While the fluid speeds are largest at early times in the simulations (Fig.~\ref{fig:SimSnaps_Samp7} and Fig.~\ref{fig:SimSnaps_Samp4}), corresponding to relatively rapid demixing, the vortex quadrupoles appear throughout the coarsening process even at late times, whenever there is a domain merger (Supplementary Movies 1 and 2). Including hydrodynamic interactions ($\gamma\neq 0$) improves the comparison between theory and experiment, as is evident by comparing Fig.~\ref{fig:AutocorrExp}(a) and~\ref{fig:AutocorrExp}(b); specifically, hydrodynamic effects tend to accelerate coarsening at the intermediate timescale of approximately 20 hours for the parameters corresponding to the BCAT-5 experiments. The fact that the experiments were conducted in microgravity made such hydrodynamic effects visible, as sedimentation was avoided and these intermediate timescales could be probed.

Our model makes a number of assumptions for the sake of simplicity. In particular, the phase-field model treats the colloid particles as a continuum field. As a consequence, we cannot treat directly the influence of colloid particles on the fluid, and instead force the fluid motion through the Korteweg stress. Moreover, we assume that the polymer chemical potential is constant, and neglect the influence of inhomogeneities in the polymer concentration. In spite of these significant simplifications, our simulation results in Fig.~\ref{fig:AutocorrExp} show that the order of magnitude predictions for the coarsening rate are correct.

Experiments have suggested that polymer redistribution is a critical ingredient driving colloidal crystallization~\cite{Palberg2009}. In the future, we could improve the model by removing our assumption that the polymer chemical potential is constant, and instead evolve the polymer concentration in time. To that end, generalized free-volume theories of the form described in Ref.~\cite{Fleer2008} could be used. These theories have the advantage that they can be extended to the regime $\xi\gtrsim 1$, which is beyond the regime of validity of the theory used in our paper (which is based on Ref.~\cite{Lekkerkerker1992}) but relevant to the BCAT-3 and BCAT-4 experiments (Appendix~\ref{Sec:Appendix}). More sophisticated models can be developed by going beyond free-volume theories, for instance by using the so-called polymer reference interaction site model (PRISM)~\cite{Fuchs2000,Fuchs2001}, which exhibits better agreement with experiment in certain regimes~\cite{Ramakrishnan2002}. 

The simulation platform detailed herein could also be extended to model more complicated colloidal systems such as suspensions of active colloids~\cite{Spellings2015}, which have attracted recent interest due to their ability to form new materials and mimic living matter~\cite{Bishop2023}. Much like the colloid-polymer systems described herein, such systems would also benefit from investigations in microgravity environments, as the destructive effects of gravitational forces, buoyancy-driven flows and hydrostatic pressure would be eliminated~\cite{Decadal2023}. External forces could also be readily added to our model~\eqref{CHStokes}, which would allow us to probe the possibility of assembling and holding colloidal structures in desired configurations. The associated protocols could potentially be realized in microgravity experiments, which could pave the way for building mobile and reconfigurable colloidal machines that manipulate their environment in controlled ways~\cite{Decadal2023}.

\begin{acknowledgements}
The authors acknowledge support from NASA grant \#80NSSC20K0274.
\end{acknowledgements}


\appendix


\section{Details of BCAT-3 and BCAT-4 experiments}\label{Sec:Appendix}

\begin{figure*}[hbtp]
    \centering
    \includegraphics[width=1\textwidth]{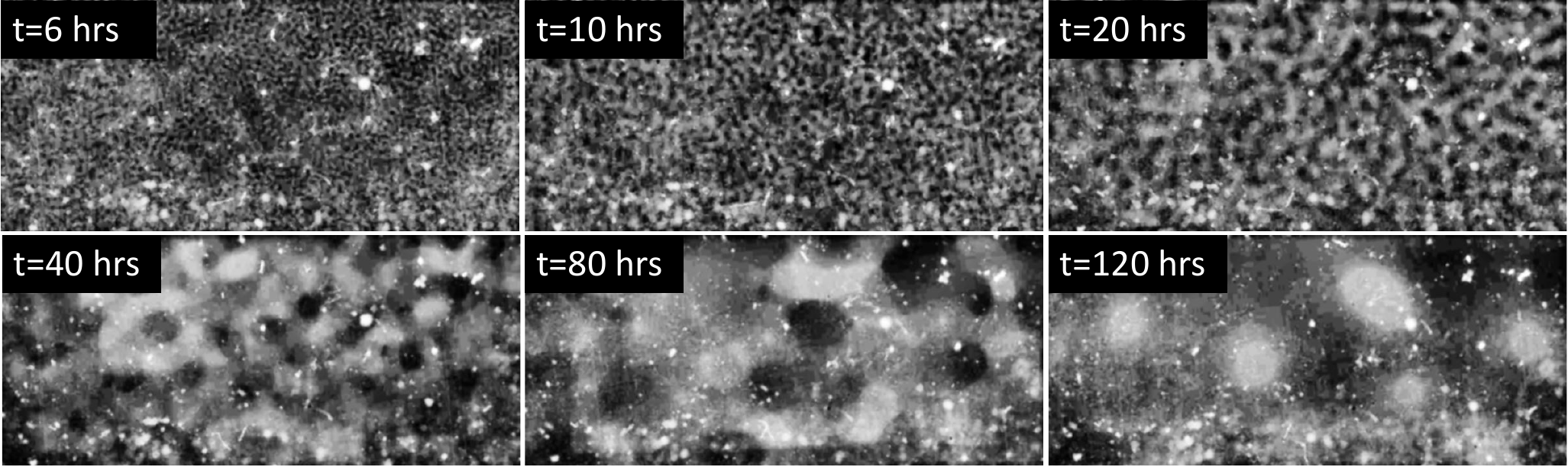}
    \caption{Time evolution of BCAT-3, Sample 1. The parameters are listed in Table~\ref{tab:param}.}
    \label{fig:BCAT3-1}
\end{figure*}

\begin{figure*}[hbtp]
    \centering
    \includegraphics[width=1\textwidth]{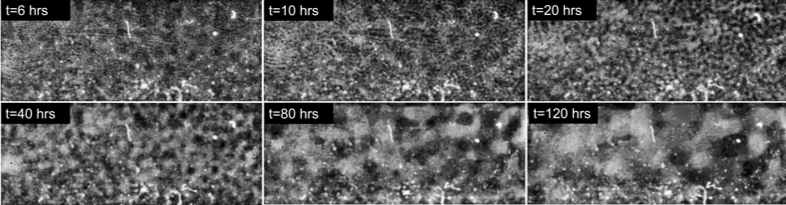}
    \caption{Time evolution of BCAT-3, Sample 2. The parameters are listed in Table~\ref{tab:param}.}
    \label{fig:BCAT3-2}
\end{figure*}

\begin{figure*}
    \centering
    \includegraphics[width=1\textwidth]{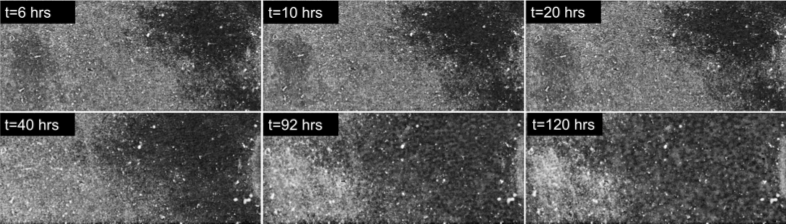}
    \caption{Time evolution of BCAT-3, Sample 4. The parameters are listed in Table~\ref{tab:param}.}
    \label{fig:BCAT3-4}
\end{figure*}

\begin{figure*}
    \centering
    \includegraphics[width=1\textwidth]{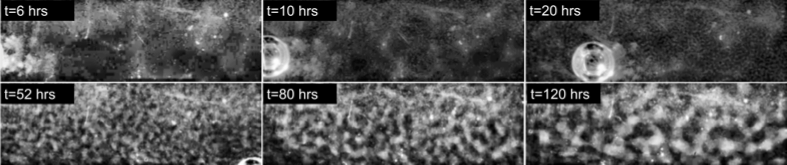}
    \caption{Time evolution of BCAT-3, Sample 6. The parameters are listed in Table~\ref{tab:param}. The circular object in the images taken at $t=10$ hours and $t=20$\ hours is the magnetic stir bar used by the astronauts to mix the samples.}
    \label{fig:BCAT3-6}
\end{figure*}

\begin{figure*}
    \centering
    \includegraphics[width=1\textwidth]{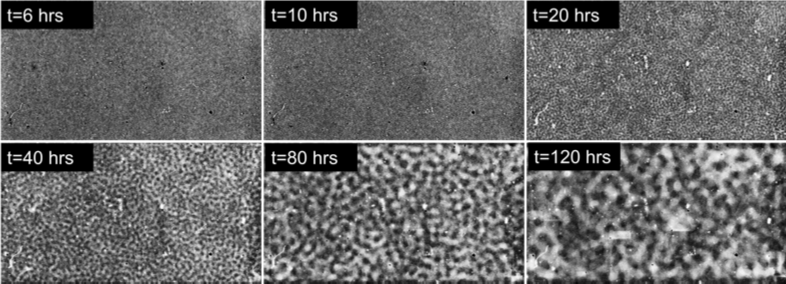}
    \caption{Time evolution of BCAT-4, Sample 2. The parameters are listed in Table~\ref{tab:param}.}
    \label{fig:BCAT4-2}
\end{figure*}

\begin{figure*}
    \centering
    \includegraphics[width=1\textwidth]{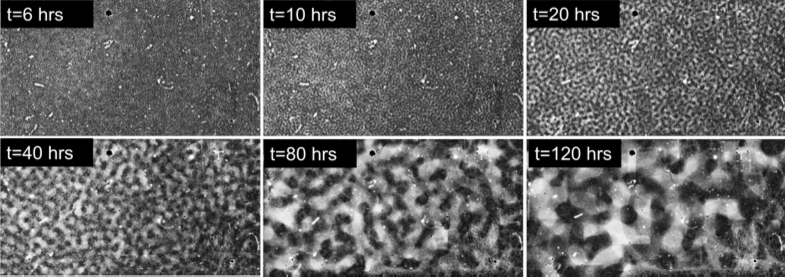}
    \caption{Time evolution of BCAT-4, Sample 3. The parameters are listed in Table~\ref{tab:param}.}
    \label{fig:BCAT4-3}
\end{figure*}

In this Appendix we collect the processed images for the BCAT-3 and BCAT-4 experiments. The parameter values are given in Table~\ref{tab:param} of the Main Text. Figures~\ref{fig:BCAT3-1}--\ref{fig:BCAT3-6} show the time evolution of the four BCAT-3 samples, and Figs.~\ref{fig:BCAT4-2}--\ref{fig:BCAT4-3} shows that for two of the BCAT-4 samples. We did not include a figure for BCAT-4 Sample 1, for which only a few images were available likely because the camera malfunctioned during the experiment.

\begin{figure*}[hbtp]
    \centering
    \includegraphics[width=1\textwidth]{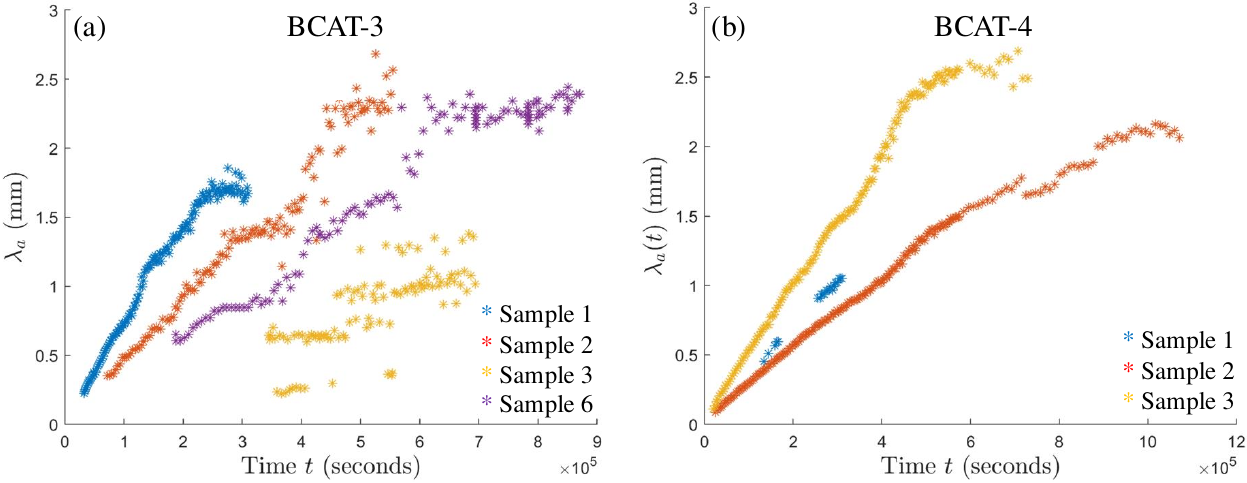}
    \caption{Time evolution of the characteristic length scale $\lambda_a(t)$ for the samples in the BCAT-3 (panel (a)) and BCAT-4 (panel (b)) experiments. Note that comparatively few images were available for BCAT-4 Sample 1, likely because the camera malfunctioned during the experiment.}
    \label{fig:BCAT34Auto}
\end{figure*}

The time evolution of the characteristic length scale $\lambda_a(t)$, as defined in \S\ref{Sec:Image} of the Main Text, is shown in Fig.~\ref{fig:BCAT34Auto}. For BCAT-3 (Fig.~\ref{fig:BCAT34Auto}(a)), it appears that higher colloid volume fractions $\phi_0$ and polymer concentrations $\rho$ are correlated with faster growth of the phase domains. Note that the colloid radius $a$ and polymer radius of gyration $\delta$ are the same for all of these samples. In addition, the values of $\phi_0$ and $\rho$ are varied together between samples: specifically, both values are largest for Sample 1, followed by Sample 2, then Sample 6 and then Sample 4. Thus, it is unclear which of the two parameters has a greater influence on the coarsening rate, which appears to increase with both $\phi_0$ and $\rho$ in Fig.~\ref{fig:BCAT34Auto}(a).

For the BCAT-4 experiment (Figure~\ref{fig:BCAT34Auto}(b)), $\phi_0$ and $\rho$ are similarly decreased together going from Samples 1 through 3, and $a$ and $\delta$ are the same for all samples (Table~\ref{tab:param} in the Main Text). Samples 1 and 2 evidently follow the same trend as the BCAT-3 samples: specifically, the coarsening rate of Sample 1 is evidently higher than that of  Sample 2, and  $\phi_0$ and $\rho$ are larger for the former. On the other hand, Sample 3 does not follow this rule, as its coarsening rate is the largest but its $\phi_0$ and $\rho$ values are the smallest. It is unclear why this is the case, making this single sample an outlier. 

\section*{Captions for Supplementary Movies}

\noindent {\bf Movie 1:} Simulation of BCAT-5 Sample 7, with $\gamma=130$; the other parameters are listed in Table~\ref{tab:param}. The left, middle and right panels correspond to the colloid volume fraction $\phi$, fluid speed $|\gamma\bs{u}|$ and fluid vorticity $\gamma\bs{\nabla}\times\bs{u}$, respectively. \\

\noindent {\bf Movie 2:} Simulation of BCAT-5 Sample 4, with $\gamma=93$; the other parameters are listed in Table~\ref{tab:param}. The left, middle and right panels correspond to the colloid volume fraction $\phi$, fluid speed $|\gamma\bs{u}|$ and fluid vorticity $\gamma\bs{\nabla}\times\bs{u}$, respectively. \\

\noindent {\bf Movie 3:} Simulations of BCAT-5 Sample 4 (top row) and BCAT-5 Sample 7 (bottom row), without (left column) and with (right column) hydrodynamics. The panels show the colloid volume fraction $\phi$. While $\gamma=0$ for the two videos in the first column, $\gamma=93$ and $\gamma=130$ for BCAT-5 Sample 4 and Sample 7, respectively.

\clearpage

 \bibliography{ColloidBib}

\end{document}